\documentclass[twocolumn, apj, numberedappendix, appendixfloats]{openjournal_dhayaa}

\usepackage{xcolor}
\usepackage{textgreek}
\usepackage[utf8]{inputenc}
\usepackage[english]{babel}

\usepackage{hyperref}
\hypersetup{
    unicode, 
    colorlinks=true,
    linkcolor=linkcolor,
    citecolor=linkcolor,
    filecolor=linkcolor,
    urlcolor=linkcolor,
}
\usepackage{xurl}
\usepackage{color,colortbl}
\definecolor{linkcolor}{rgb}{0.0,0.3,0.5}
\usepackage{tensind}
\tensordelimiter{?}
\DeclareGraphicsExtensions{.bmp,.png,.jpg,.pdf}
\usepackage{verbatim}
\usepackage[normalem]{ulem}
\usepackage{orcidlink}
\usepackage{soul}
\usepackage{lineno}
\renewcommand{\AA}{%
  \leavevmode\setbox0=\hbox{A}\copy0\kern-\wd0
  \raise1.3ex\hbox to\wd0{\hss{\fontsize{6}{6}\selectfont$\circ$}\hss\kern.01em\kern.04em}%
}
\linenumbers

\begin{document}
\title{The DECam MAGIC Survey-- Mapping the Ancient Galaxy in CaHK: Overview and Summary of Early Science}

\shorttitle{MAGIC Survey Overview \& Early Science}


\author{A.~Chiti\,$^\dagger$\orcidlink{0000-0002-7155-679X}}
\affil{Kavli Institute for Particle Astrophysics \& Cosmology, Stanford University, Stanford, CA 94305, USA}
\affil{Brinson Prize Fellow}

\author{A.~Drlica-Wagner\orcidlink{0000-0001-8251-933X}}
\affil{Fermi National Accelerator Laboratory, P.O.\ Box 500, Batavia, IL 60510, USA}
\affil{Kavli Institute for Cosmological Physics, University of Chicago, Chicago, IL 60637, USA}
\affil{Department of Astronomy \& Astrophysics, University of Chicago, Chicago, IL 60637, USA}
\affil{NSF-Simons AI Institute for the Sky (SkAI), 172 E. Chestnut St., Chicago, IL 60611, USA}

\author{A.~B.~Pace\orcidlink{0000-0002-6021-8760}}
\affil{Department of Astronomy, University of Virginia, 530 McCormick Road, Charlottesville, VA 22904, USA}

\author{W.~Cerny\orcidlink{0000-0003-1697-7062}}
\affil{Department of Astronomy, Yale University, New Haven, CT 06520, USA}

\author{K.~R.~Atzberger\orcidlink{0000-0001-9649-8103}}
\affil{Department of Astronomy, University of Virginia, 530 McCormick Road, Charlottesville, VA 22904, USA}

\author{F.~O.~Barbosa\orcidlink{0000-0002-8262-2246}}
\affil{Universidade de S\~ao Paulo, Instituto de Astronomia, Geof\'isica e Ci\^encias Atmosf\'ericas, Departamento de Astronomia, SP 05508-090, S\~ao Paulo, Brasil}

\author{J.~A.~Carballo-Bello\orcidlink{0000-0002-3690-105X}}
\affil{Instituto de Alta Investigaci\'on, Universidad de Tarapac\'a, Casilla 7D, Arica, Chile}

\author{H.~Q.~Do\orcidlink{0009-0009-3497-8369}}
\affil{Department of Astronomy \& Astrophysics, University of Chicago, Chicago, IL 60637, USA}

\author{A.~P.~Ji\orcidlink{0000-0002-4863-8842}}
\affil{Kavli Institute for Cosmological Physics, University of Chicago, Chicago, IL 60637, USA}
\affil{Department of Astronomy \& Astrophysics, University of Chicago, Chicago, IL 60637, USA}
\affil{NSF-Simons AI Institute for the Sky (SkAI), 172 E. Chestnut St., Chicago, IL 60611, USA}

\author{G.~Limberg\orcidlink{0000-0002-9269-8287}}
\affil{Kavli Institute for Cosmological Physics, University of Chicago, Chicago, IL 60637, USA}
\affil{Department of Astronomy \& Astrophysics, University of Chicago, Chicago, IL 60637, USA}

\author{A.~M.~Luna\orcidlink{0009-0009-9570-0715}}
\affil{Department of Astronomy \& Astrophysics, University of Chicago, Chicago, IL 60637, USA}

\author{C.~E.~Mart\'inez-V\'azquez\orcidlink{0000-0002-9144-7726}}
\affil{NSF NOIRLab, 670 N. A'ohoku Place, Hilo, Hawai'i, 96720, USA}

\author{V.~M.~Placco\orcidlink{0000-0003-4479-1265}}
\affil{NSF NOIRLab, 950 N. Cherry Ave., Tucson, AZ 85719, USA}

\author{D.~S.~Prabhu\orcidlink{0000-0002-8217-5626}}
\affil{Department of Astronomy/Steward Observatory, 933 North Cherry Avenue, Room N204, Tucson, AZ 85721-0065, USA}

\author{G.~S.~Stringfellow\orcidlink{0000-0003-1479-3059}}
\affil{University of Colorado Boulder, Boulder, CO 80309, USA}

\author{A.~K.~Vivas\orcidlink{0000-0003-4341-6172}}
\affil{Cerro Tololo Inter-American Observatory/NSF NOIRLab, Casilla 603, La Serena, Chile}

\author{A.~R.~Walker\orcidlink{0000-0002-7123-8943}}
\affil{Cerro Tololo Inter-American Observatory/NSF NOIRLab, Casilla 603, La Serena, Chile}

\author{S.~N.~Campana}
\affil{Department of Physics and Astronomy, Dartmouth College, Hanover, NH 03755, USA}

\author{J.~L.~Carlin\orcidlink{0000-0002-3936-9628}}
\affiliation{Rubin Observatory/AURA, 950 North Cherry Avenue, Tucson, AZ, 85719, USA}

\author{V.~Chandra\orcidlink{0000-0002-0572-8012}}
\affil{Center for Astrophysics $|$ Harvard \& Smithsonian, 60 Garden Street, Cambridge, MA 02138, USA}

\author{D.~Crnojevi\'c\orcidlink{0000-0002-1763-4128}}
\affil{Department of Physics \& Astronomy, University of Tampa, 401 West Kennedy Boulevard, Tampa, FL 33606, USA}

\author{P.~S.~Ferguson\orcidlink{0000-0001-6957-1627}}
\affil{DiRAC Institute, Department of Astronomy, University of Washington, 3910 15th Ave NE, Seattle, WA, 98195, USA}

\author{J.~J.~Hermes\orcidlink{0000-0001-5941-2286}}
\affil{Department of Astronomy, Boston University, 725 Commonwealth Avenue, Boston, MA 02215, USA}

\author{N.~Kallivayalil\orcidlink{0000-0002-3204-1742}}
\affil{Department of Astronomy, University of Virginia, 530 McCormick Road, Charlottesville, VA 22904, USA}

\author{G.~E.~Medina\orcidlink{0000-0003-0105-9576}}
\affil{David A. Dunlap Department of Astronomy \& Astrophysics, University of Toronto, 50 St George Street, Toronto ON M5S 3H4, Canada}
\affil{Dunlap Institute for Astronomy \& Astrophysics, University of Toronto, 50 St George Street, Toronto, ON M5S 3H4, Canada}

\author{M. Navabi\orcidlink{0000-0001-9438-5228}}
\affil{Department of Physics, University of Surrey, Guildford GU2 7XH, Surrey, UK}

\author{N.~E.~D.~Noël\orcidlink{0000-0002-8282-469X}}
\affil{Department of Physics, University of Surrey, Guildford, GU2 7XH}

\author{A.~H.~Riley\orcidlink{0000-0001-5805-5766}}
\affil{Lund Observatory, Division of Astrophysics, Department of Physics, Lund University, SE-221 00 Lund, Sweden}

\author{D.~J.~Sand\orcidlink{0000-0003-4102-380X}}
\affil{Department of Astronomy/Steward Observatory, 933 North Cherry Avenue, Room N204, Tucson, AZ 85721-0065, USA}

\author{C.~W.~Skeffington}
\affil{University of Colorado Anschutz, Aurora, CO 80045, USA}

\author{K.~Tavangar\orcidlink{0000-0001-6584-6144}}
\affil{Department of Astronomy, Columbia University, New York, NY 10027, USA}

\author{J.~F.~Wu\orcidlink{0000-0002-5077-881X}}
\affil{Space Telescope Science Institute, 3700 San Martin Drive, Baltimore, MD 21218, USA}
\affil{Center for Astrophysical Sciences, Department of Physics \& Astronomy, Johns Hopkins University, Baltimore, MD 21218, USA}

\author{Y.~Choi\orcidlink{0000-0003-1680-1884}}
\affil{NSF NOIRLab, 950 N. Cherry Ave., Tucson, AZ 85719, USA}

\author{D.~Erkal\orcidlink{0000-0002-8448-5505}}
\affil{School of Mathematics and Physics, University of Surrey, Guildford GU2 7XH, UK}

\author{D.~J.~James\orcidlink{0000-0001-5160-4486}}
\affil{ASTRAVEO LLC, PO Box 1668, Gloucester, MA 01931, USA}
\affil{Applied Materials Inc., 35 Dory Road, Gloucester, MA 01930, USA}

\author{T.~S.~Li\orcidlink{0000-0002-9110-6163}}
\affil{David A. Dunlap Department of Astronomy \& Astrophysics, University of Toronto, 50 St George Street, Toronto ON M5S 3H4, Canada}

\author{P.~Massana\orcidlink{0000-0002-8093-7471}}
\affil{NSF NOIRLab, Casilla 603, La Serena, Chile}

\author{B.~Mutlu-Pakdil\orcidlink{0000-0001-9649-4815}}
\affil{Department of Physics and Astronomy, Dartmouth College, Hanover, NH 03755, USA}

\author{D.~L.~Nidever\orcidlink{0000-0002-1793-3689}}
\affil{Department of Physics, Montana State University, P.O. Box 173840, Bozeman, MT 59717-3840, USA}

\author{K.~A.~G.~Olsen\orcidlink{0000-0002-7134-8296}}
\affil{NSF NOIRLab, 950 N. Cherry Ave., Tucson, AZ 85719, USA}

\author{J.~D.~Sakowska\orcidlink{0000-0002-1594-1466}}
\affil{Instituto de Astrofísica de Andalucía (CSIC), Glorieta de la Astronomía,  E-18080 Granada, Spain}

\author{L.~Santana-Silva\orcidlink{0000-0003-3402-6164}}
\affil{Observatório do Valongo/UFRJ, Ladeira do Pedro Antônio, 43 - Centro, Rio de Janeiro - RJ, 20080-090, Brazil}

\author{J.~D.~Simon\orcidlink{0000-0002-4733-4994}}
\affil{Observatories of the Carnegie Institution for Science, 813 Santa Barbara Street, Pasadena, CA 91101, USA}

\author{E.~Tollerud\orcidlink{0000-0002-9599-310X}}
\affil{Space Telescope Science Institute, 3700 San Martin Drive, Baltimore, MD 21218, USA}

\author{A.~Zenteno\orcidlink{0000-0001-6455-9135}}
\affil{Cerro Tololo Inter-American Observatory/NSF NOIRLab, Casilla 603, La Serena, Chile}


\collaboration{MAGIC \& DELVE Collaborations}

\email{$^\dagger$ achiti@stanford.edu}

\begin{abstract}

We present the DECam Mapping the Ancient Galaxy in CaHK (MAGIC) survey, a 54-night NOIRLab Survey Program to image $\gtrsim$5,000\,$\deg^2$ of the southern hemisphere using a metallicity-sensitive narrow-band filter covering the Ca\,{\sc{ii}}~H\&K lines centered at 3955\,{\AA}. 
This filter is installed on the Dark Energy Camera (DECam), mounted on the 4-m NSF V\'{i}ctor M. Blanco Telescope.  
The survey reaches typical $10\sigma$ depths of $\text{mag}_{\text{CaHK}} \approx 22.5$, 3--4\,mag deeper than comparable surveys in the southern hemisphere.
By combining photometry from this Ca~{\sc{ii}}~H\&K filter with existing DECam $g,r,i$ broadband photometry from the DECam Local Volume Exploration (DELVE) survey, MAGIC is deriving photometric metallicities for red giant branch stars down to the magnitude limit of usable proper motions from \textit{Gaia} data release 3 (DR3). 
MAGIC has already imaged $\sim$3{,}000\,$\deg^2$, supplemented by other affiliated observing programs that have used this filter to image star clusters, dwarf galaxies, and stellar streams. 
We overview MAGIC's survey strategy, describe data processing through the derivation of metallicities and photometric distances, and summarize early science results that have been published with this dataset.
In addition, we present several new results, including the confirmation of a distant ($>5\,r_h$) member of the Reticulum II ultra-faint dwarf galaxy, on-sky density maps of low-metallicity stars into the distant Milky Way halo ($\sim150$\,kpc) recovering 13/14 ultra-faint dwarf galaxies in the current footprint, and a validation of our initial targeting of extremely metal-poor stars.
Collectively, these results demonstrate that the MAGIC dataset enables cutting-edge studies of the faint, low-metallicity regime of the Milky Way and its substructures. 
    
\end{abstract}


\maketitle

\section{Introduction} 

The local universe is an active discovery space, facilitating studies of the Milky Way's assembly \citep[e.g.,][]{ivezic2012, bee+2013, blandhawthorn2016, a+20, deasonbelokurov2024}, early galaxy formation \citep[e.g.,][]{btg+14,s+19}, the small-scale distribution of dark matter \citep[e.g.,][]{bb+17, bp+25}, and the production of elements across cosmic time \citep[e.g.,][]{nomoto2013,fn+15,kkl+20}. 
For instance, wide-field photometric surveys such as the Sloan Digital Sky Survey (SDSS; \citealt{yaa+00}) enabled the discovery of ultra-faint dwarf galaxies (UFDs; \citealt{wdm+05, zbe+06}), which are the most dark-matter-dominated and faintest known galaxies (mass-to-light ratios M/L$ \gtrsim 100$\,M$_\odot$/L$_\odot$; $L \lesssim 10^5$\,L$_\odot$; \citealt{sg+07, mdr+08, mcs+18, s+19}). 
SDSS and other such surveys (e.g., Pan-STARRS, DES, DECaLS, DELVE, \textit{Gaia}; \citealt{cmm+16, des1, Dey:2019, dcn+21, gaia1, gaiadr3}) have additionally revealed a large population of stellar streams around the Milky Way (e.g., \citealt{bze+06, gd+06, sdb+18, ljp+22}), which trace its gravitational potential and are sensitive to perturbations from starless dark matter halos (e.g., \citealt{ebb+16}).
The \textit{Gaia} space mission \citep{gaia1, gaiadr3} has now measured proper motions for over a billion stars, providing evidence for the last major merger of the Milky Way \citep[known as Gaia-Sausage/Enceladus;][]{bee+18, haywood+2018, hbk+18} and motions on the plane of the sky for nearly every known Milky Way satellite galaxy \citep{mv+20, lhb+21,pel+22}.
Complementing these studies, spectroscopic surveys (e.g., SEGUE, LAMOST, GALAH, APOGEE, S5, DESI; \citealt{yrn+09,zzc+12,dfb+15,msf+17, lkz+19,desi}) have collectively provided information on the chemical composition of millions of stars.

Stars in the Milky Way halo and its faint substructures (e.g., dwarf galaxies and stellar streams) are generally metal-poor\footnote{``Metal-poor" is formally defined as having [Fe/H] $< -1.0$, where [Fe/H] = $\log_{10}(N_{\rm Fe}/N_{\rm H})_{\star} 
- \log_{10}(N_{\rm Fe}/N_{\rm H})_\sun$ \citep{bc+05, f+18}}.
Low-metallicity stars, where ``metallicity" is defined as the abundance of elements heavier than helium, function as time capsules for understanding the high-redshift universe \citep{bc+05, fn+15}.
Their stellar photospheres preserve the chemical composition of their natal gas cloud (barring certain identifiable circumstances, such as binary mass transfer or evolution on the red giant branch; e.g., \citealt{gsc+00, ltb+05, h+05, hansen2016cempno, hansen2016cemps}), meaning that low-mass, long-lived stars with low-metallicities reflect the composition of the interstellar medium when it was enriched by relatively few supernovae.
Accordingly, the relative elemental composition of stars as a function of metallicity can trace element production over time. 
Stars at the lowest metallicities, i.e., extremely metal-poor stars (EMPs; [Fe/H] $< -3.0$), or ultra metal-poor stars (UMPs; [Fe/H] $< -4.0$), are hypothesized to be enriched by so few events that they provide insight on the nucleosynthetic yields of the first stars \citep{hw+02,cds+04, fac+05,cbf+11,kbf+14,jfb+15,pfb+16,itk+18,hym+18,vss+23}. 

This has motivated searches for the lowest metallicity stars spanning several decades \citep[e.g.,][]{bond1970, b+80, bidelman1973, bps+85, c+03, christlieb2008, ksb+07, beb2014, smy+17}, driven both by finding low-mass, metal-free first stars (aka. ``Population III") and by inferring the properties (e.g., initial masses, explosion energies) of the first generation of metal-producing supernovae through their nucleosynthetic yields \citep{tun+07, hw+10, pfl+15, svs+24}. 
Surveys for low-metallicity stars in the Milky Way historically obtained low-resolution prism/grism spectroscopy, typically leveraging the Ca~{\sc{ii}}~H\&K lines at 3968.5\,{\AA} and 3933.7\,{\AA}.
These features were targeted since they remain prominent even in the lowest metallicity stars observed with low resolution ($R\lesssim1500$) spectroscopy, making the identification of stars with weak absorption features feasible (e.g., Figure 5 in \citealt{paa+22}). 

Recent untargeted searches for low-metallicity stars have leveraged photometric metallicities \citep[e.g.,][]{ksb+07, smy+17,whitten2019, wpb+21,galarza2022, fzw+23}, building on the long-recognized fact that blanketing at bluer wavelengths by metal lines, and particularly absorption from the prominent Ca~{\sc{ii}}~H\&K lines can lead to a mapping between blue intermediate- and narrow-band filters and stellar metallicity \citep[e.g.,][]{sn+89, alp+89}. 
Such work has shown that this photometry can provide quantitative measurements of the stellar metallicity down to [Fe/H] $\approx -3.0$ \citep[e.g.,][]{isj+08,abj+13,cfm+21,hbw+22,msy+24,bcl+25} and enables the identification of the lowest metallicity stars (i.e., with [Fe/H] $< -4.0$; \citealt{kbf+14, sab+18, nbd+19, lmj+21, prl+21, plc+25, cml+24, cpp+25, limberg2025}).
Collectively, these efforts have discovered ${\sim}50$ known stars in the [Fe/H] $< -4.0$ regime \citep{sky+08,af+18,slm+19,bcf+25}, directly showing these stars tend to be exceptionally carbon-enhanced \citep[e.g.,][]{pfb+14, apl+22} leading to a range of theories on how the first stars may have enriched primordial dark matter halos \citep[e.g.,][]{iut+05,mem+06,cm+14,hy+19,vss+23}.

\begin{figure*}[htbp]
    \centering
    \includegraphics[width=1\textwidth]{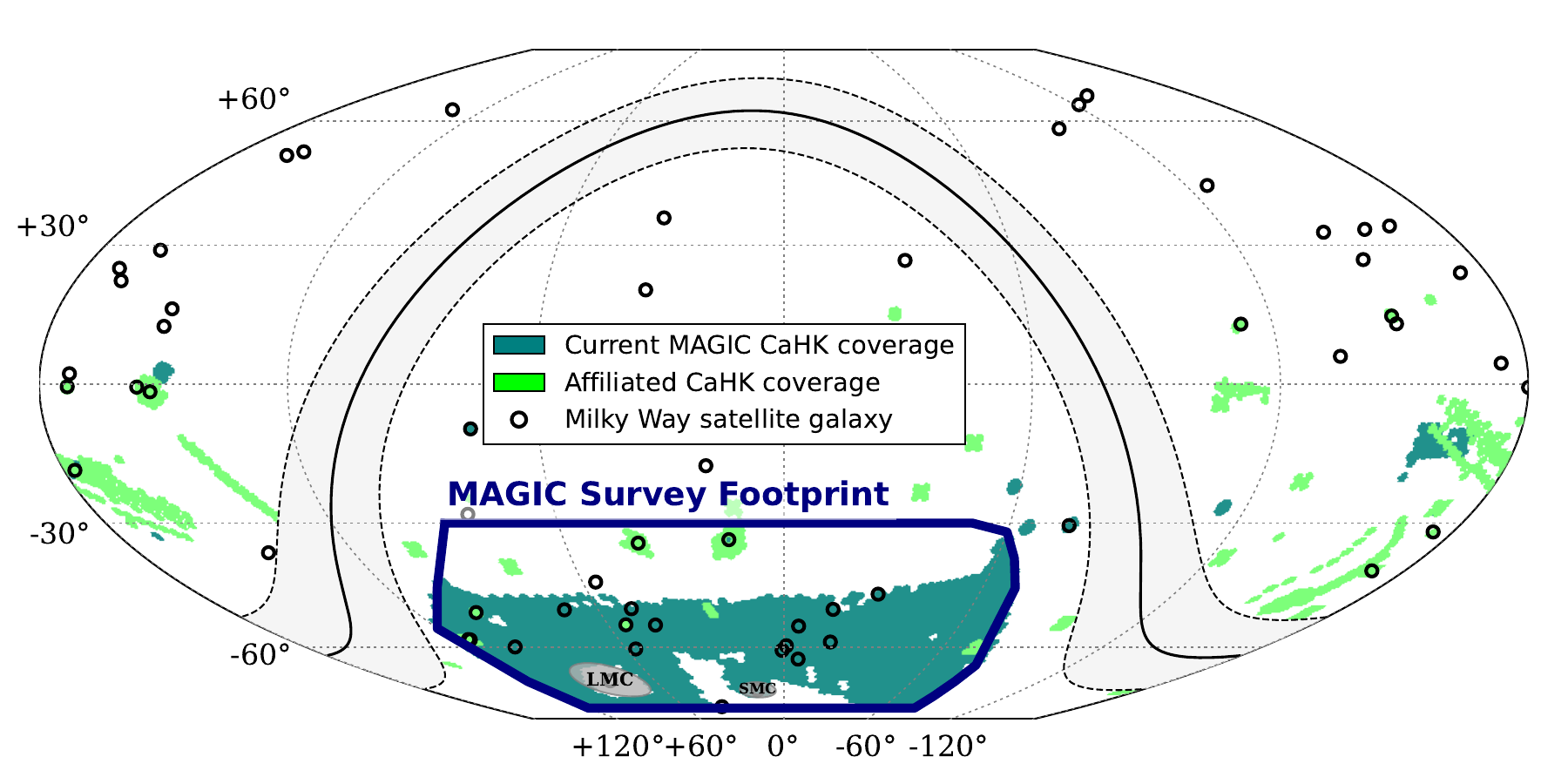}
    \caption{An equal-area sky map using the McBryde-Thomas flat polar quartic projection \citep{McBryde:1949} in equatorial coordinates (right ascension, declination) showing the planned DECam MAGIC survey footprint (outlined in dark blue). Existing coverage from MAGIC (teal) and affiliated DECam CaHK P.I.\ programs (green) are shown. Confirmed dwarf galaxies from the Local Volume Database \citep{pace+25} are overplotted as hollow black circles, and the MCs are shown as grey ellipses. The Galactic plane is overplotted as a solid line, with regions at low Galactic latitude ($|b| < 10^\circ$) shaded in gray between dotted lines.
    The primary targets of the affiliated P.I. programs are listed in Table~\ref{tab:affiliated} in Appendix~\ref{app:affiliated}.}
    \label{fig:footprint}
\end{figure*}

A key benefit of photometric metallicity surveys is that they simultaneously provide metallicities for all stars that are detected with sufficient signal-to-noise (S/N) in the field-of-view of the imager, in contrast to fiber-fed/multi-slit spectroscopic surveys which have a spatial selection function or suffer from incompleteness. 
This characteristic makes photometric metallicity surveys a critical complement to wide-field broad-band and astrometric imaging surveys (e.g., SDSS, DES, \textit{Gaia}).
For instance, recent work has shown that a combination of photometric metallicities, proper motion, and color-magnitude diagram (CMD) selection can provide insight on the formation and dynamical state of UFDs by detecting extended features in their peripheries \citep[][]{cfs+21,lja+22,ljb+23,ocs+24,pcd+25,cpp+25}.
Similarly, wide-field photometric metallicities have been used to confirm the existence of faint stellar streams in the Milky Way halo through their coherent metallicities \citep[][]{mis+22}, including one with a metallicity below the apparent floor for globular clusters (GCs; \citealt{mva+22}), in addition to faint structures in the vicinity of GCs \citep[e.g.,][]{kik+25}.
Large datasets of metallicities within dwarf galaxies can also be used to trace their star formation histories and evolution (e.g., \citealt{tih+04,kls+11,fws+23, fws+24, bcl+25, hkt+25}); similarly, metallicity distribution measurements across the Milky Way halo can uncover its assembly history \citep[e.g.,][]{dkf+23}, constrain the properties of Population III star enrichment \citep[e.g.,][]{hmc+22}, and study the formation of the inner galaxy \citep[e.g.,][]{rca+22}.
Such datasets can also enable comparative studies of the enrichment of satellites associated with the LMC \citep[e.g.,][]{pkg+20, pel+22}, which may have formed in a distinct early environment from Milky Way satellite galaxies.

These datasets also enable the discovery of rare objects (e.g., low-metallicity stars) across environments (i.e., dwarf galaxies, streams) with well-described photometric selection functions as targets for current and next-generation multiplexed surveys (4MOST, Subaru-PFS, DESI, WEAVE, Spec-S5; \citealt{4most, tts+16, desi, weave, specs5}).
These facets motivate a wide-field, photometric survey that provides metallicities into the extremely low metallicity regime (i.e., [Fe/H] $\lesssim -3.0$) reaching depths ($g\approx20.5$) complementing the latest astrometric datasets (i.e., \textit{Gaia} DR3; \citealt{gaiadr3}).
These synergies would naturally facilitate broad science across the Milky Way halo.

Here we provide an overview for the DECam Mapping the Ancient Galaxy in CaHK (MAGIC; NOIRLab Prop. ID 2023B-646244; PI: Anirudh Chiti) survey, designed to derive photometric metallicities for stars across a quarter of the southern hemisphere (${\sim}5{,}000$\,deg$^2$) extending $\gtrsim 3$ magnitudes fainter than existing narrow-band photometric metallicities in this region of the sky (SkyMapper, SPLUS, \textit{Gaia} DR3 XP; \citealt{ksb+07, mrs+19, gaiaxp, 2024A&A...691A.138P}; see Section~\ref{sec:processing}). 
MAGIC uses the Dark Energy Camera \citep[DECam;][]{fdh+15} on NSF's 4-m V\'{i}ctor M. Blanco Telescope at Cerro Tololo Inter-American Observatory (CTIO), Chile to image the sky, using a newly installed narrow-band filter covering the Ca~{\sc{ii}}~H\&K lines (hereafter CaHK filter\footnote{Although formally denoted "N395" by CTIO and in the NOIRLab Astro Data Archive.}; central wavelength 3955\,{\AA}, 101\,{\AA}-wide)\footnote{\url{https://noirlab.edu/science/programs/ctio/filters/Dark-Energy-Camera}}, modeled after the filter used in the Pristine survey in the Northern hemisphere \citep{smy+17}. 
Notably, the ${\sim}$5,000\,deg$^2$ MAGIC footprint covers the Magellanic Clouds (MCs) and a large number of dwarf galaxies discovered within the Dark Energy Survey \citep[DES;][]{Bechtol:2015, Koposov:2015, Kim:2015, Drlica-Wagner:2015}, enabling a near-comprehensive chemical mapping of the Milky Way halo and its prominent satellite systems (see Figure~\ref{fig:footprint}).
These observations are complemented by targeted P.I.\ programs led by MAGIC collaborators that have obtained CaHK data covering stellar streams, GCs, UFDs, and classical dwarf spheroidal (dSph) galaxies over much of the southern hemisphere (e.g., \citealt{apk+26, dcf+26}).

The paper is organized as follows. 
In Section~\ref{sec:survey}, we overview the survey aims and planned footprint, data processing, and existing observations; in Section~\ref{sec:photometric_metallicities}, we summarize the photometric performance, metallicity derivation, and validate the quality of the metallicities in our catalog; in Section~\ref{sec:science}, we summarize early science results, and present several new results; in Section~\ref{sec:conclusion}, we conclude and discuss upcoming data releases. 

\section{Survey aims, current observations, and data processing}
\label{sec:survey}

The DECam MAGIC survey is designed to enable four key scientific goals: (1) To discover and enable spectroscopic follow-up of the lowest metallicity stars ([Fe/H] $< -4.0$); (2) To characterize the metallicity distribution function (MDF) and spatial metallicity gradients of the Milky Way and its dwarf satellite galaxies; (3) To detect low-surface-brightness features in the Milky Way environment; and (4) To probe spatially extended populations around the Milky Way's metal-poor satellite systems. 
The dataset can also be used to facilitate a number of additional science cases, including probing the star formation histories of the MCs, generating metallicity-resolved maps in the vicinity of the wake of the Large Magellanic Cloud (LMC), the serendipitous discovery of quenched dwarf galaxies through the 4000\,{\AA} break, the identification of quasars at redshift $z\approx2.2$, and the discovery of metal-polluted white dwarfs.

The primary scientific goals of MAGIC, as elaborated on in Section~\ref{sec:aims},  require the determination of reliable photometric metallicities. 
In Sections~\ref{sec:strategy} and~\ref{sec:processing}, we outline the MAGIC survey data collection and our data processing procedure in order to produce catalogs of astronomical objects. 
We note that in Section~\ref{sec:photometric_metallicities}, we discuss the photometric metallicity determination and validation.
Broadband photometric data are required to derive metallicities and achieve any of the auxiliary scientific goals of the survey. 
Accordingly, we cross-match our source catalogs with DECam $g,r,i$ photometry provided by the second data release of the DECam Local Volume Exploration survey \citep[DELVE DR2;][]{delvedr2}\footnote{\url{https://delve-survey.github.io}}\textsuperscript{,}\footnote{\url{https://datalab.noirlab.edu/data/delve}}.
Sources in MAGIC data releases will also be provided with astrometry from the \textit{Gaia} DR3 \citep{gaia1, gaiadr3}, when available.

\subsection{Scientific aims}
\label{sec:aims}


\textbf{(1) Discovery and follow-up of second-generation stars:} 
Stars with [Fe/H] $< -4.0$ (and plausibly some with [Fe/H] $< -3.0$) have likely been enriched by a single generation of supernovae, meaning that they preserve direct signatures from  Population III stars (\citealt{fn+15}). 
However, existing data is sparse due to the rarity of the most metal-poor stars; to date, only ${\sim}50$ stars are known with [Fe/H] $< -4.0$ \citep[e.g.,][]{slm+19}. 
Nonetheless, we see hints of critical insights from this small sample; for instance, nearly all stars in this regime (${\sim}$80\%) are enhanced in carbon, and very few are deficient in [Mg/Ca] \citep{sky+08, pfb+14, apl+22}. 
This can affect the shape of the initial mass function (via cooling mechanisms) of the second and subsequent generations of stars, and the mass distribution of Population III stars themselves \citep[e.g.,][]{chiaki2017}. 
While observations of the MDF of the Milky Way halo have conflicting trends at the extremely metal-poor regime (e.g., \citealt{scc+09, ynb+13, dbm+19, ysm+21, cmf+21}), \citet{bbj+19} presented a calculation assuming the MDF follows the power law presented in \citet{scc+09}, indicating a rate of 1 star with [Fe/H] $< -4.0$ per ${\sim}100$\,deg$^2$ down to a brightness of $V = 18$. Nominally, this calculation indicates that the MAGIC footprint may enclose ${\sim}$50 stars above that brightness limit with [Fe/H] $< -4.0$, suitable for detailed chemical abundance studies with high-resolution spectroscopy.
Based on current compilations (e.g., \citealt{slm+19}), ${\sim}10$ known UMP stars lie in the MAGIC survey footprint. 

The MAGIC footprint will also provide meaningful insights on how the above properties vary across early environments, including in the MCs and other satellites of the Milky Way. 
Only a handful of stars in the LMC with [Fe/H] $< -2.5$ have recently had detailed chemical abundances published \citep{ond+24, cml+24, jcm+25}. 
Given that the LMC fell into the Milky Way only ${\sim}$2\,Gyr ago (e.g., \citealt{bkh+07, kvb+13, gbl+19, conroy2021}; although see \citealt{v+24}), its lowest metallicity stars carry the signatures of the first supernova in a formerly different region of the universe \citep{cml+24}. 
This makes it an ideal test for variations in the properties of the first stars across environments, and the universality of early enrichment.
Initial results from MAGIC have already provided a detailed chemical abundance analysis of a star with [Fe/H] $< -4.0$ in the distant Milky Way halo \citep{plc+25}, and have presented the detection and analysis of the first known star in this regime (with [Fe/H] $< -4.6$) in an ultra-faint dwarf galaxy \citep{cpp+25}.

\textbf{(2) The MDF of the Milky Way and its satellites:}
Recent models of the Milky Way MDF show that the low-metallicity tail ([Fe/H] $\lesssim -3.0$) is sensitive to parameters such as the initial mass function of Population III stars, early star-formation efficiency, and early feedback prescriptions \citep{tarumi2020, ksy+20, hartwig2022}. 
The DECam MAGIC survey will allow us to derive metallicities for $\mathcal{O}$($10^3$) stars with [Fe/H] $< -3.0$.
These measurements will form model constraints on the initial mass function of the first stars \citep[e.g., whether the maximum possible mass of Population III stars includes or excludes pair instability supernovae;][]{tarumi2020}, the star formation efficiency of Population III star-forming minihalos \citep{debennassuti2017}, and the stochasticity of feedback from Population III stars \citep{hartwig2022}. 
The MAGIC dataset will also strongly constrain the overall MDF, which is sensitive to the assembly history of the Milky Way \citep{dkf+23}. 

Apart from the detailed MDFs extending to [Fe/H] $ \lesssim -3.0$, the new photometric metallicities will also allow homogeneous analyses of metallicity gradients and other complex chemo-dynamics (e.g., asymmetric gradients, signatures of mergers) inside dwarf galaxies within the footprint of the survey. Similar exercises in the literature usually require inhomogeneous compilations of [Fe/H] values from spectroscopy and are limited in radius around these Milky Way satellites \citep{taibi2022}.
Therefore, the wide and homogeneous coverage of MAGIC, as well as the expected precision of our derived metallicities (see Section~\ref{sec:validation}), carries the potential for exploring the driving mechanisms of stellar-population variations including dwarf-dwarf mergers, in-situ feedback, or other external mechanisms.
An initial result from the MAGIC survey dataset has already derived a spatially complete (out to ${\sim}7\,r_h$) metallicity map of the Sculptor dSph, demonstrating a break in the metallicity gradient \citep[See Section~\ref{sec:lit};][]{bcl+25}.


\textbf{(3) Detecting low-metallicity substructures in the Milky Way and MCs:}
The Milky Way's low-metallicity substructures (e.g., accreted dwarf galaxies, stellar streams) and the complex substructures in the MCs link to several fundamental questions, including the nature of dark matter and the assembly history of our Galaxy. 
The dozens of metal-poor stellar streams in the Milky Way halo provide a test for $\Lambda$CDM, as they are sensitive to perturbations from its predicted population of low-mass (M$_{\text{vir}}$ $< 10^8$ M$_\odot$) dark matter halos \citep{jsh+02, ili+02, ebb+16} and also trace the mass distribution of the Galaxy \citep{jzs+99, ili+01, koposov2023}. 
Substructure in the periphery of the MCs probes the dynamical history of these satellite galaxies, which in turn influence the Milky Way halo through their induced dark matter wake \citep{garavito2021, dsb+26}. 
The discovery of such streams and substructure is still ongoing, accelerated by \textit{Gaia} astrometry \citep{mi+18, mis+22, imt+24}.

Stellar streams and substructures are nearly all at low-metallicities ($\langle$[Fe/H]$\rangle < -1.0)$ and are obscured by metal-rich Milky Way foreground stars, meaning that the CaHK filter presents a unique opportunity for significant foreground removal in substructure searches to access this low surface brightness regime. 
For instance, the Pristine survey has confirmed the existence of many stellar streams through the coherent metallicities of their stars, in addition to detecting several more candidate streams \citep{mis+22}. 
MAGIC will enable such studies over a contiguous footprint in the southern hemisphere with deeper narrow-band photometry (see Section~\ref{sec:processing}), in a particularly promising region for faint substructure searches due to a high density of known dwarf galaxies and several stellar streams. 
This complements existing DES stream searches at low surface brightnesses that span part of our footprint \citep{sdb+18}.
Early results from the program have already detected the presence of faint (${\sim}35$--36\,mag\,arcsec$^{-2}$) tidal tails around the Crater II dwarf galaxy \citep{apk+26}.

The MAGIC survey also has the potential to provide the largest and deepest metallicity-resolved map of the outer halo to ${\sim}200$~kpc. 
Metallicities are the dominant uncertainty in determining isochrone distances to outer halo giants \citep{conroy2021}, and including metallicity information from the survey can reduce the distance uncertainty from ${\sim}50\%$ to ${\sim}10\%$ at $100$~kpc and ${\sim}30\%$ at $200$~kpc. 
This will enable precise measurements of the outer halo's stellar density profile, and given our survey's wide coverage, clues about its shape and dynamical response to perturbations from LMC infall \citep{conroy2021,garavito2021}.

\textbf{(4) Probing extended features around faint Milky Way satellites, star clusters, and stellar streams:}
The Milky Way's UFDs are analogs of some of the earliest systems that formed in the universe \citep{s+19}, being largely composed of very metal-poor stars ($\langle$[Fe/H]$\rangle \lesssim -2.0$) and are the most dark-matter dominated systems known (M/L $> 100$\,M$_\odot$/L$_\odot$; e.g., \citealt{sg+07, mdr+08, gpp+26, g+26}). 
The outskirts of UFDs are increasingly recognized as promising areas to probe signatures of tidal disruption \citep{li2018}, the possible extents of their dark matter halos \citep{cfs+21, dbf+22, jhs+24}, or evidence of early galactic mergers \citep{tarumi2021, qpb+25}. 
This discovery space was previously obscured due to the faintness of dwarf outskirts \citep[${\sim}33$\,mag\,arcsec$^{-2}$;][]{rey2019}, but recent work has demonstrated that a combination of \textit{Gaia} proper motions and photometric metallicities can uncover stars in these low surface brightness outskirts \citep[e.g,][]{cfs+21, nja+22, ljb+23, pcd+25} for targeted spectroscopic follow-up \citep[e.g.,][]{chiti2023, wvs+23, szv+23, ocs+24}. 

These approaches can also readily be extended to the Milky Way's GCs to uncover faint features in their outskirts \citep[e.g., ][]{kik+25}.
These regions have generated interest due to searches for signatures of the tidal disruption of GCs \citep[e.g.,][]{rog+02, s+20, bpp+20, zmd+22, ckp+25}, associations with stellar streams \cite[e.g.,][]{imt+24}, or the presence of diffuse outskirts that may reflect their progenitor \citep[][]{bif+03, csm+14, kns+22}. 
The same sensitivity to faint, low-metallicity features can be used to trace morphological features in stellar streams.
These include cocoons, gaps, spurs, and off-track overdensities \citep[e.g., ][]{mic+19, bhp+19, bpp+20} that can trace heating by dark matter subhalo impacts, dynamical interactions with nearby systems, or other components of the Milky Way \citep[e.g., ][]{ili+02, jsh+02, ebb+16, ppj+17, ebl+19, kbl+19, lke+21, nbl+24}. 
Photometric metallicities can be used to both detect these features by removing foreground metal-rich stars, and also confirm features by matching their metallicities to those of a host stream.

MAGIC's survey design will enable such studies across its footprint, covering a significant fraction of known dwarf galaxies (see Figure~\ref{fig:footprint}) and a number of star clusters and streams. This will allow population-level constraints on the frequency (or lack) of these features across the UFD population, in addition to the chemical abundances and kinematics of stars sufficiently bright for spectroscopy. The kinematics of these stars will be used to assess their dynamical state via tests for velocity gradients \citep{li2018}, and their chemical abundances (e.g., iron-peak, $\alpha$-elements) can possibly disentangle the mechanisms of early stellar halo formation (see Section~\ref{sec:retii}).
Demonstrating the promise of the survey for such work, targeting from MAGIC data has already enabled the discovery of a star in the Pictor II UFD outskirts ($>5\,r_h$) that clearly hosts an enrichment signature from Population III supernovae \citep{cpp+25}.
Similarly, MAGIC metallicities have also already helped characterize the morphology of the Jet stellar stream, confirming a clear diffusion on one end \citep{dcf+26}.


\subsection{Data Collection \& Observing Strategy}
\label{sec:strategy}

\begin{figure}[t!]
    \centering
    \includegraphics[width=1\columnwidth]{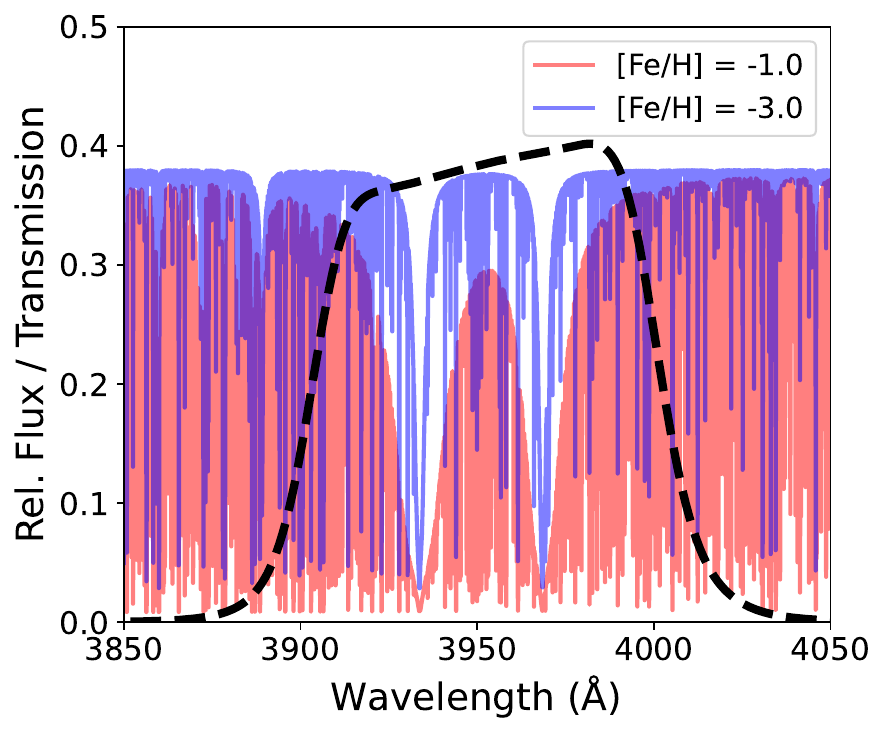}
    \caption{The bandpass of the DECam CaHK filter, convolved with the telescope and mirror response, is shown as a dashed black line.
    Synthetic stellar spectra at [Fe/H]~=~$-1.0$ and [Fe/H]~=~$-3.0$ and fixed $T_{\text{eff}}=5000\,$K, $\log\,g = 2.0$\,dex over-plotted. The variation in metallicity appreciably changes the strength of the Ca\,{\sc ii} H \& K lines, which alters the observed flux through the narrow-band filter.}
    \label{fig:tophat}
\end{figure}

The DECam MAGIC survey utilizes the ${\sim}3$\,deg$^2$ field-of-view DECam imager on the 4\,m Blanco Telescope located at CTIO \citep{fdh+15}.
A narrowband filter centered on the Ca\,{\sc ii}~H\&K lines was purchased by NOIRLab for community use in 2022.
The filter was manufactured by Asahi Spectra Co.\ Ltd., the vendor that fabricated most of the DECam filters \citep{faa+10}, using the sputtering process, in a facility specifically designed to produce very large and high quality astronomical filters.
The CaHK filter passband was deliberately chosen to be nearly identical to the filter used for the Pristine survey on CFHT \citep{smy+17}. The as-built filter has a central wavelength (CWL) of 3951.90\,{\AA} and a full width at half maximum (FWHM) of 100.10\,{\AA}, with the passband shape that is close to ``top hat'' (see Figure~\ref{fig:tophat}), having a nearly flat transmission of ${\sim}$95\,\% for $\pm 35$\,{\AA} to either side of the CWL.
The transmission uniformity is maintained to $\pm1.25\%$ (peak-to-valley) over the full field.
The remaining filter characteristics were chosen to be specific for DECam, namely a circular substrate of fused silica with diameter 620\,mm and thickness 14\,mm, with coatings on both sides.
Particular care was taken to ensure that the out-of-band blocking is adequate, given the very high red sensitivity of the DECam CCDs.
Further details are available on the CTIO website, or by contacting CTIO. 

The DECam MAGIC survey was allocated 54 nights to cover a region of ${\sim}5,000$\,deg$^2$ that largely overlaps with DES (Figure~\ref{fig:footprint}). 
Observations began in August 2023, and the survey has collected 1192 science-quality exposures (1294 total exposures) as of the end of the 2025B observing semester (January 2026). 
Over this period,  41.5 nights of data have been allocated through the survey program, from which ${\sim}$25 nights of usable data have been obtained with the remainder lost due to weather and mechanical losses. 
In addition, members of the MAGIC collaboration have led multiple programs in collaboration with MAGIC.
These include NOIRLab Prop.\ IDs: 2023A-933926 (P.I. Chiti), 2024A-167177 (P.I. Cerny), 2024A-930400 (P.I. Cerny), 2024A-974884 (P.I. Pace), 2024B-255918 (P.I. Pace), 2024B-616175 (P.I. Cerny), 2025A-303722 (P.I. Carballo-Bello, Chiti), 2025A-402104 (P.I. Pace), 2025A-568024 (P.I. Do, Chiti), 2025A-942172 (P.I. Mart\'inez-V\'azquez), 2025B-444868 (P.I. Pace), for a total of 20.5 allocated nights (see Table~\ref{tab:affiliated} in Appendix~\ref{app:affiliated}).
These programs largely target objects of interest (e.g., star clusters, dwarf galaxies, stellar streams) and comprise the majority of the DECam CaHK data collected outside the MAGIC survey footprint (Figure~\ref{fig:footprint}).

A significant amount of infrastructure exists to facilitate observations from DECam, including the SISPI data acquisition and control interface \citep{haa+08}, the \texttt{kentools} package to inventory and inspect images, the DECam Community Pipeline for data reduction \citep{vg+14}, and the NOIRLab Astro Archive to host processed DECam frames\footnote{\url{https://astroarchive.noirlab.edu}}.
MAGIC observations are coordinated through the umbrella of the DELVE Collaboration, with the majority of observing performed remotely.
We note that each MAGIC image has a proprietary period of one year, becoming publicly available for download on the NOIRLab Astro Archive at that point after data collection. 

Motivated by our scientific goals to be able to discriminate metal-poor/metal-rich stars down to the limit of usable \textit{Gaia} proper motions, we nominally forecasted a desired S/N $\approx20$ at $g\approx20.5$ (see Section~\ref{sec:photometric_metallicities} for details).
Based on pilot data with the DECam CaHK filter collected on the nights of 2023 February 21--22, we forecasted that this would be achievable in 12\,minutes in good seeing conditions.
Given that shallower CaHK data already exists in the southern hemisphere (e.g., S-PLUS, or from the \textit{Gaia} XP catalog; \citealt{mrs+19, gaiaxp}), we optimized for increased depth in MAGIC. 
In regions of high interest in our footprint (e.g., dwarf galaxies), we obtain three dithered twelve-minute pointings to fill in chip gaps and minimize loss from cosmic rays.

\begin{figure*}[htbp]
    \centering
    \includegraphics[width=0.33\textwidth]{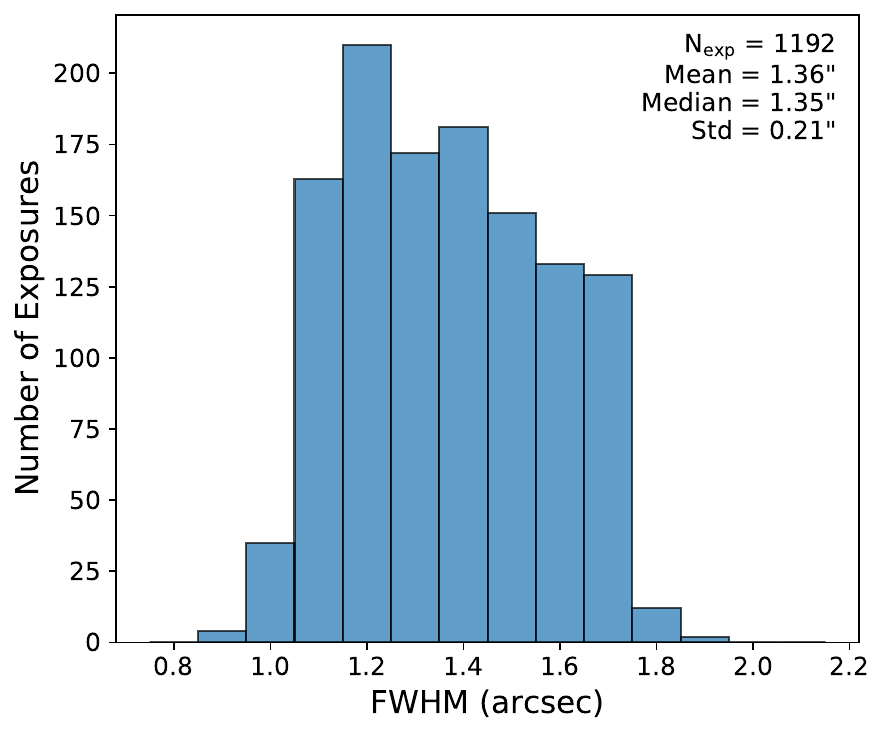}
    \includegraphics[width=0.33\textwidth]{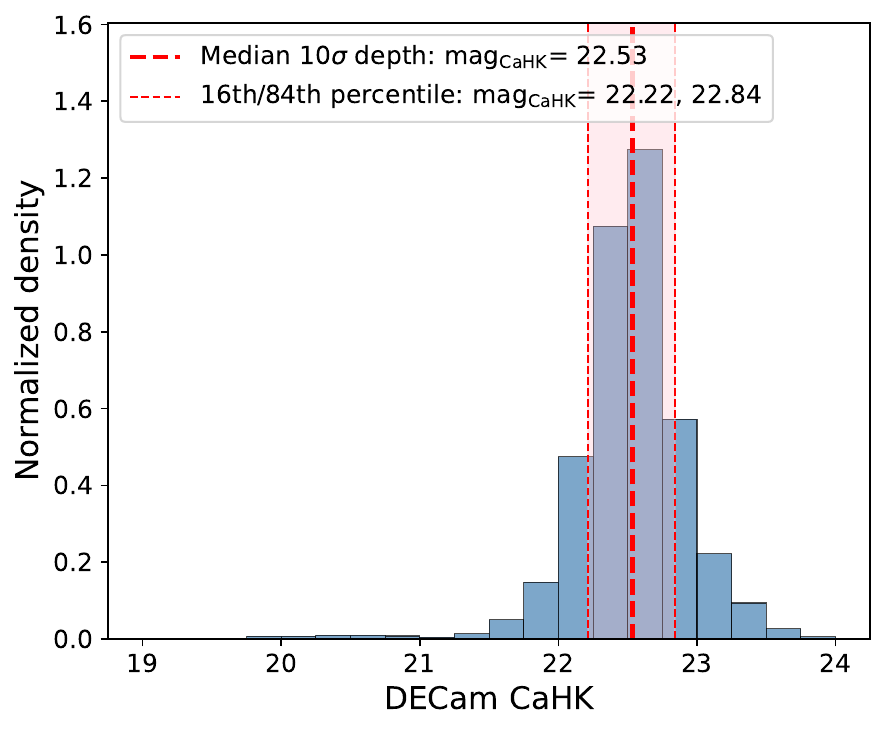}
    \includegraphics[width=0.33\textwidth]{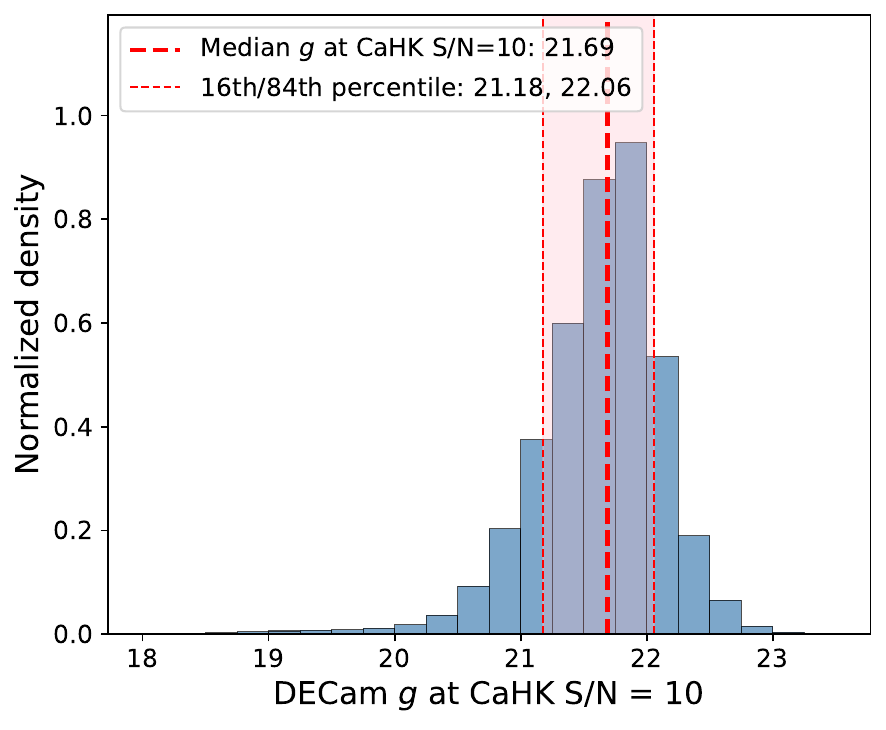}
    \caption{Left: A histogram of the FWHM of the PSF in DECam CaHK exposures obtained for the MAGIC Survey. This information is taken from image metadata from the NOIRLab Astro Data Archive, and ought to correspond to the typical FWHM of sources across the CCD mosaic.
    Middle \& Right: Histograms of the median 10\,$\sigma$ depth in each DECam CaHK image processed for the MAGIC survey, based on the PSF-fitted magnitudes derived from \texttt{PSFEx} (see Section~\ref{sec:processing}).
    The middle panel shows the 10$\sigma$ depth in CaHK magnitudes, and the right panel shows the corresponding DELVE DR2 $g$ magnitudes of sources at the 10$\sigma$ CaHK detection limit.}
    \label{fig:psfs_depths}
\end{figure*}

For observations, we generate DECam pointings to cover the footprint shown in Figure~\ref{fig:footprint} using the scheme adopted for the DELVE survey \citep{delvedr2}, which is derived from the same all-sky icosahedral tiling scheme used by DECaLS \citep{Dey:2019}. 
The specific pointings were selected using the \texttt{obztak} scheduler\footnote{\url{https://github.com/kadrlica/obztak}} updated to include a polygon for the MAGIC footprint. 
This approach generated 1913 DECam pointings that span the entirety of the ${\sim}$5,000\,deg$^2$ footprint, with a ${\sim}20$\,\% overlap between adjacent pointings.
We generate the observing schedule for each night in a given semester by manually adjusting weights in \texttt{obztak}, based on scientific priorities (e.g., coverage of dwarf galaxies, contiguous regions, and the MCs).
We then dynamically update plans for subsequent nights as data is collected.
Within each night, we monitor data quality by taking a short (30\,s) $g$-band exposure every hour and deriving its effective-exposure-time scale factor ($t_{\text{eff}}$; a proxy for depth defined in \citealt{nbg+16}) using the in-built \texttt{qcInv} command in the DECam observer terminal.
We exclude from further processing exposures that are very shallow (i.e., meta-column ``depth" $<$ 21.5\,mag in the NOIRLab Astro Data Archive), or corrupted images (e.g., where guiding failed), and re-image those tiles as needed. 
A distribution of the FWHM values of the point spread function (PSF) in our survey images is shown in Figure~\ref{fig:psfs_depths}.

\subsection{Processing \& Catalog Generation}
\label{sec:processing}

We retrieve DECam CaHK images for our program from the NOIRLab Astro Data Archive, which hosts raw images in addition to images that have been reduced using the DECam Community Pipeline (CP; \citealt{vg+14}).
Specifically, we retrieve single-epoch images for each pointing, along with their associated weight-maps and data-quality masks generated by the CP. 
These images are photometered using the standard \texttt{Source Extractor} and \texttt{PSFEx} software \citep{ba+96, b+11}, with configuration files from the DES Data Management Pipeline \citep{Desai2012, Mohr2012, mgm+18, desdr2}.
The outcome of this procedure is a catalog with standard \texttt{Source Extractor/PSFEx} keys, including instrumental magnitudes, photometric uncertainties, and flags indicating sources with bad pixels in their photometered region\footnote{\url{https://sextractor.readthedocs.io/en/latest/Flagging.html}}.
We note that photometry from DECam CCDs N30 and S7, at the edges of the focal plane, is excluded due to known issues\footnote{\url{https://noirlab.edu/science/programs/ctio/instruments/Dark-Energy-Camera/Known-Problems}}.

We then cross-match the resulting source catalogs from each exposure with the publicly available DELVE DR2 for broadband DECam $g, r, i$ photometry \citep{delvedr2}, in addition to \textit{Gaia} DR3 \citep{gaiadr3} using the \texttt{cdsskymatch} command in the \texttt{stilts}\footnote{\url{https://www.star.bris.ac.uk/~mbt/stilts/}} software \citep{stilts}.
In the resulting catalogs, we require sources to be detected in DELVE DR2 to be retained, given the greater depth of the broadband photometry, but retain sources without entries in \textit{Gaia} DR3 due to the deeper source detection depth in the DECam CaHK images.

We perform a photometric zero-point calibration for the CaHK source catalog from each pointing using spectro-photometric data from the \textit{Gaia} XP spectra \citep{gaiaxp, gaiaxp2}, in a similar process as \citet[][also \citealt{2024A&A...691A.138P}]{msy+24}.
Briefly, \textit{Gaia} DR3 released low-resolution ($R\approx25-100$) flux-calibrated spectra of ${\sim}$220 million sources from the BP and RP spectrophotometers (hereafter, XP spectra). 
Additionally, the mission provided functionality in the \texttt{GaiaXPy}\footnote{\url{https://gaia-dpci.github.io/GaiaXPy-website/}}\citep{gaiaxpy} Python library to derive synthetic photometry for sources from XP spectra in photometric bandpasses of interest, including a narrow-band CaHK filter with analogous properties to DECam N395 (i.e., based on the CaHK filter used in the Pristine survey; \citealt{smy+17})\footnote{However, we note that the quantum efficiency of the DECam CCDs at ${\sim}3950$\,{\AA} differs from that of the MegaCam CCDs used for Pristine.}. 
We compute synthetic CaHK magnitudes for sources with \textit{Gaia} XP spectra, and then cross-match the catalogs from each DECam CaHK pointing to this global catalog to derive a photometric zero-point correction for each pointing.
These corrections are computed after excluding sources with synthetic photometric CaHK uncertainties $> 0.05$\,mag, and after excluding galaxies (i.e., \texttt{SPREAD\_MODEL} $> 0.003$ from \texttt{PSFEx} for $g \lesssim 19$; e.g., \citealt{Desai2012}) and sources with bad pixels  (e.g., cosmic rays, bad pixels, saturation; \texttt{FLAGS} $> 3$, \texttt{IMAFLAGS\_ISO} $> 0$, \texttt{SPREAD\_MODEL} $< -0.003$, in \texttt{Source Extractor}).
These cross-matches typically have ${\sim}160$ sources for zero-point calibration per pointing, with a tail of ${\sim}$2\% of exposures having no valid sources for zero-point calibration. 
Future processing will incorporate other techniques, such as UberCal \citep{psf+08}, to extend zero-point corrections to these exposures. 

After the photometric zero-point calibration, the individual source catalogs are merged to form the final catalog. 
Multiple photometric measurements of individual sources are averaged together using inverse-variance weighting, with the standard error on the inverse-variance weighted mean adopted as the final uncertainty. 
Suspect photometry from individual epochs are excluded via \texttt{FLAGS} $> 3$ from the \texttt{Source Extractor} (i.e., indicating saturation) or \texttt{IMAFLAGS\_ISO}$> 0$ (i.e., bad pixels/cosmic ray affecting source photometry). 
The final CaHK magnitudes are de-reddened using the reddening coefficient $A_{\text{CaHK}}/E(B-V)_{\text{SFD98}} = 3.924$ and the dustmap from \citet{sfd+98}, following \citet{smy+17}, which holds as we have performed a zero-point of our photometry to be on its magnitude scale. 
In the middle and right panels of Figure~\ref{fig:psfs_depths}, we show the resulting 10$\sigma$ depths of the CaHK filter in our images as a function of CaHK magnitude, and the $g$-band magnitude distribution of sources that have S/N = 10 in CaHK.
These indicate a typical 10$\sigma$ depth of mag$_{\text{CaHK}} = 22.5$ in our survey footprint, with a median $g=21.69$ for sources at this depth in CaHK.
For comparison, SPLUS DR2 lists a typical 10\,$\sigma$ depth in their CaHK filter (J0395) of 19\,mag (Figure 23 in \citealt{ash+22}), SkyMapper DR4 lists a typical 10\,$\sigma$ depth in their $v$ filter of 18.9\,mag \citep{owb+24}, and narrow-band CaHK photometry from \textit{Gaia} XP reaches S/N $\approx 10$ at mag$_{\text{CaHK}}\approx$18\,mag (from the dataset in \citealt{msy+24}). 
The CaHK photometry from the Pristine survey has S/N = 10 at a typical $G\approx20.0$ ($g\approx$ 20.5) for sources with $BP-RP \approx 1.0$ \citep{msy+24}; the equivalent in the MAGIC catalog is $g\approx21.7$.

\section{Deriving Photometric Metallicities, Distances \& Validation}
\label{sec:photometric_metallicities}

\begin{figure*}[htbp]
    \centering
    \includegraphics[width=1\textwidth]{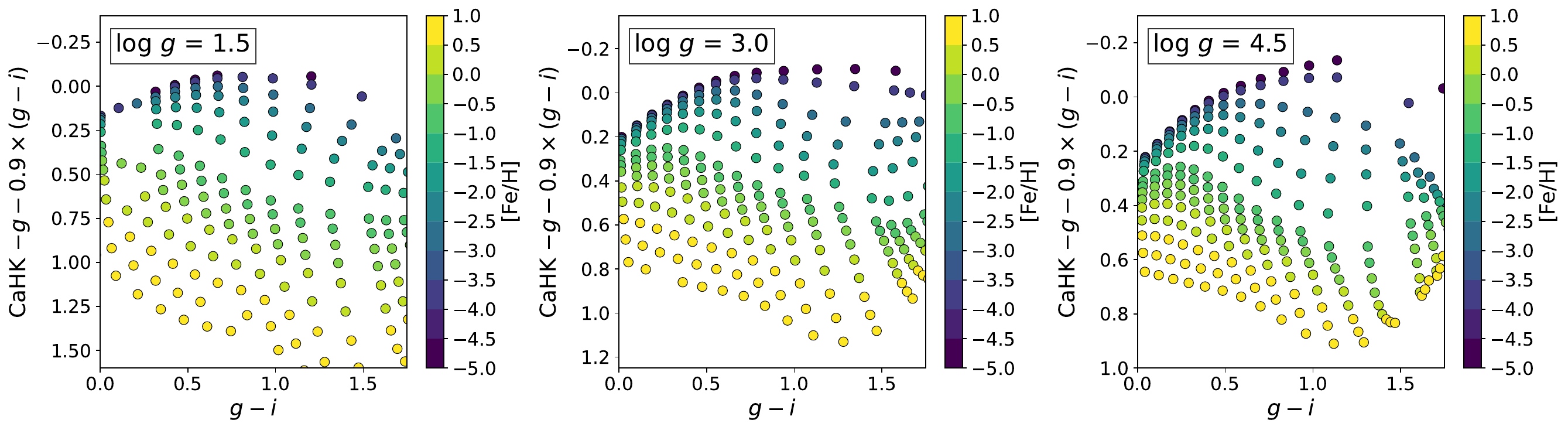}
    \caption{Left: Color-color plot including the CaHK photometry on the y-axis, generated using the grid of synthetic photometry detailed in Section~\ref{sec:grid}. Stars at different metallicities separate cleanly in this color-color space, enabling a mapping from CaHK $- g - 0.9\times(g-i)$ and $g-i$ to [Fe/H]. Middle \& Right: Same as left, but at different surface gravities ($\log\,g$). Note that the grids of synthetic photometry shift as surface gravity varies.}
    \label{fig:synthfeh}
\end{figure*}

In this section, we describe our methods for deriving photometric metallicities from the source catalogs described in Section~\ref{sec:processing}. 
Briefly, we derive metallicities by comparing our observed CaHK photometry, along with broadband DECam $g$ and $i$ photometry, to a grid of forward-modeled synthetic photometry that predicts the magnitudes that stars ought to have across a range of stellar parameters ($T_{\text{eff}}$, $\log\,g$, [Fe/H]).
This grid allows a mapping between observed colors and metallicity, especially in color-color spaces with metallicity-sensitive magnitudes (i.e., CaHK; see Figure~\ref{fig:synthfeh}). 
Our particular implementation of this approach largely follows \citet{cfj+20, cfm+21}, which applied this technique on SkyMapper photometry \citep{ksb+07, owb+19}.
Furthermore, the analysis described here has already been implemented and tested in \citet{bcl+25}, which provided metallicities for ${\sim}3800$ stars in the Sculptor dSph, and used in target selection for various published spectroscopic follow-up studies from MAGIC imaging \citep{plc+25, cpp+25}. 

We describe the process of generating the grid of synthetic photometry and deriving photometric metallicities in Section~\ref{sec:grid} and Section~\ref{sec:derivationfeh}. 
When deriving metallicities from this approach, we have to assume a surface gravity ($\log\,g$) for each star.
We discuss how $\log\,g$ is inferred, and outline how these are mapped to photometric distances in Section~\ref{sec:derivationdistances}. 
In Section~\ref{sec:failures}, we discuss how we flag common failure modes and contaminants in the photometric metallicity catalog.
We validate our methods in Section~\ref{sec:validation}.

\subsection{Grid of synthetic photometry}
\label{sec:grid}

In order to generate our grid of synthetic photometry, we first create a grid of flux-calibrated spectra using the Turbospectrum radiative transfer code\footnote{\url{https://github.com/bertrandplez/Turbospectrum\_NLTE}} \citep{ap+98,p+12}, along with MARCS model atmospheres\footnote{\url{https://marcs.astro.uu.se}} \citep{gee+08}, and a linelist from the Vienna Atomic Line Database\footnote{\url{https://vald.astro.uu.se}} (VALD; \citealt{pkr+95, rpk+15}). 
Specifically, we use standard composition MARCS models, which assume [$\alpha$/Fe] $=0.4$ when [Fe/H] $< -1.0$ and linearly declines to [$\alpha$/Fe] = 0 at [Fe/H] = 0. 
We adopt spherical MARCS models when $\log\,g \leq 3.5$\,dex, and otherwise use plane-parallel models to generate 4490 synthetic spectra spanning $\log\,g = -0.5$ to $\log\,g = +5.0$, $T_{\text{eff}} = 2500\,$K to $T_{\text{eff}} = 8000\,$K, [Fe/H] = $-5.0$ to [Fe/H] = $+1.0$.
We note that these spectra were generated under the assumption of 1D local thermodynamic equilibrium, with a microturbulence set to 2\,km\,s$^{-1}$ for all spectra. 
Since the purpose of this grid is to forward-model photometry and not perform detailed modeling of line profiles, we opt for a fixed microturbulence to streamline analysis. 

From this grid of synthetic spectra, we derive predicted photometry in the CaHK filter and DECam $g, r, i$ filters using the formalism described in \citet{cv+14}, following the implementation in \citet{cfj+20}.
We download the system throughput for the DECam $g, r,$ and $i$ band photometry from the NOIRLab website.
For the CaHK filter, we assume the tophat function for the filter used in the Pristine survey \citep{smy+17}, which closely approximates the as-built DECam filter profile (Section~\ref{sec:strategy}) and is the bandpass to which our photometric zeropoints are calibrated in Section~\ref{sec:processing}. 
We apply an empirical offset of +0.12\,mag to the CaHK photometry in this synthetic grid, analogous to the $+0.06$\,mag offset in SkyMapper $v$ in \citet{cfj+20}, to account for model imperfections and bring the resulting photometric metallicity zero-points closer to agreement with reference values (see Section~\ref{sec:validation}).
We show color-color plots of CaHK $- g - 0.9\times(g-i)$ vs. $(g-i)$ using this grid of synthetic photometry in Figure~\ref{fig:synthfeh}, which is a canonical color-combination that is known to separate metal-rich and metal-poor stars in the literature \cite[e.g.][]{ksb+07, smy+17, cfj+20}.

\subsection{Deriving photometric metallicities}
\label{sec:derivationfeh}

We derive photometric metallicities using the aforementioned grid of synthetic photometry, largely following \citet{cfj+20, cfm+21}. 
Aspects of our methodology are also briefly described in \citet{bcl+25}, which presents a pilot MAGIC study of the Sculptor dSph.
As noted, Figure~\ref{fig:synthfeh} shows metallicity-sensitive color-color plots that are populated by entries from our grid of synthetic photometry at various $\log\,g$ values, demonstrating that a mapping is feasible between color terms and [Fe/H], at fixed input $\log\,g$. 
Accordingly, we start by using the \texttt{scipy.interpolate.griddata} python module with a cubic spline interpolation scheme to generate a two-dimensional function at each $\log\,g$ in our grid that maps the location of stars in CaHK $- g - 0.9\times(g-i)$ vs. $(g-i)$ color-color space to metallicity. 
These interpolation functions are generated separately for each $\log\,g$, due to the visible shifts in color-color space as $\log\,g$ varies (see Figure~\ref{fig:synthfeh}). 

Accordingly, a key requirement for inferring the metallicity in this framework is an estimate of the $\log\,g$ for each star. 
Given that the focus of MAGIC is to investigate the ancient components of the Milky Way (e.g., the stellar halo), we approximate $\log\,g$ by matching the observed $(g-i)_0$ color of each star to theoretical 12\,Gyr isochrones from the Dartmouth database at [Fe/H]~=~$-2.5, -2.0, -1.5$, and $-1.0$ \citep{dcj+08}\footnote{Although, $\log\,g$ values from the [Fe/H]~=~$-1.0$ isochrone are currently excluded when assigning final metallicities; see next paragraph.}, with the upper bound set by the motivation for MAGIC to target metal-poor Milky Way halo stars and lower bound set by the isochrone library.
For each isochrone, we derive two possible $\log\,g$ values from matching the $(g-i)_0$ color: one assuming the star lies on the main sequence (MS), and one assuming it lies on the red giant branch (RGB). 
We do not explicitly address stars not in these evolutionary states (e.g., horizontal branch stars, variable stars), since aspects of these sources, such as the high temperatures of blue horizontal branch stars and the necessity of time-domain information for variable stars, render photometric metallicity determinations much more complex. 

For each star, we compute metallicities for each of the MS and RGB assumptions across all four isochrones, yielding eight initial metallicity estimates. 
We then average the RGB-based metallicities and the MS-based metallicities separately. 
To select the most appropriate isochrone for the final metallicity determination, we identify which reference isochrone has a metallicity closest to the averaged value for each evolutionary state. 
The metallicity derived using the $\log\,g$ from that isochrone is adopted as the final metallicity for the MS and RGB cases. 
In current processing, we exclude $\log\,g$ values from the [Fe/H] = $-1.0$ isochrone and instead adopt those from the [Fe/H] = $-1.5$ isochrone for the metal-rich end, as the inclusion of the former introduces a metallicity systematic relative to APOGEE (see Section~\ref{sec:validation}).    
This iterative approach accounts for the coupled dependence of metallicity and $\log\,g$, i.e. that the isochrone-derived $\log\,g$ depends on assumed metallicity, and the synthetic-grid metallicity depends on assumed $\log\,g$.
The outcome of this procedure is two metallicity estimates per star: one assuming the star is on the RGB ([Fe/H]$_{\text{RGB}}$), and one assuming it is on the MS ([Fe/H]$_{\text{MS}}$).
We describe how we discriminate between these two scenarios in the next subsection.

Uncertainties on the photometric metallicities are derived by propagating the photometric uncertainties in CaHK, $g$, and $i$-band magnitudes. 
We recompute the metallicity after varying each input magnitude by its  $1\sigma$ uncertainty in the photometric catalog, and then derive a total random uncertainty by summing in quadrature the three resulting shifts in [Fe/H]. 
We then add a systematic uncertainty floor of 0.16\,dex in quadrature to account for e.g., imperfections in the synthetic grid, following \citet{cfj+20}.
This metallicity uncertainty floor also produces errors that are consistent with spectroscopic studies of red giant stars in the Sculptor dSph \citep{bcl+25}.

\subsection{Deriving photometric distances}
\label{sec:derivationdistances}

The metallicity derivation described in the previous subsection yields two estimates per star ([Fe/H]$_{\text{RGB}}$, [Fe/H]$_{\text{MS}}$). 
For a subset of these stars, we can determine their evolutionary state using the fact that the isochrone-matching procedure implicitly provides a distance estimate under each assumption (i.e., $d_{\text{RGB}}$, $d_{\text{MS}}$). 
Specifically, matching the $(g-i)_0$ color of a star to an isochrone yields an inferred absolute magnitude for the star, from which a distance modulus can be derived.
Since RGB stars are intrinsically more luminous than MS stars, the inferred $d_{\text{RGB}}$ is typically far larger than $d_{\text{MS}}$ for a star at a given apparent magnitude. 
Accordingly, the distances of these stars that are inferred from all-sky \textit{Gaia} DR3 parallax measurements \citep{gaiadr3} can be used to discriminate between the RGB and MS scenarios for a subset of sources when $g \lesssim 19.5$.

\begin{figure}[htbp]
    \centering
    \includegraphics[width=1\columnwidth]{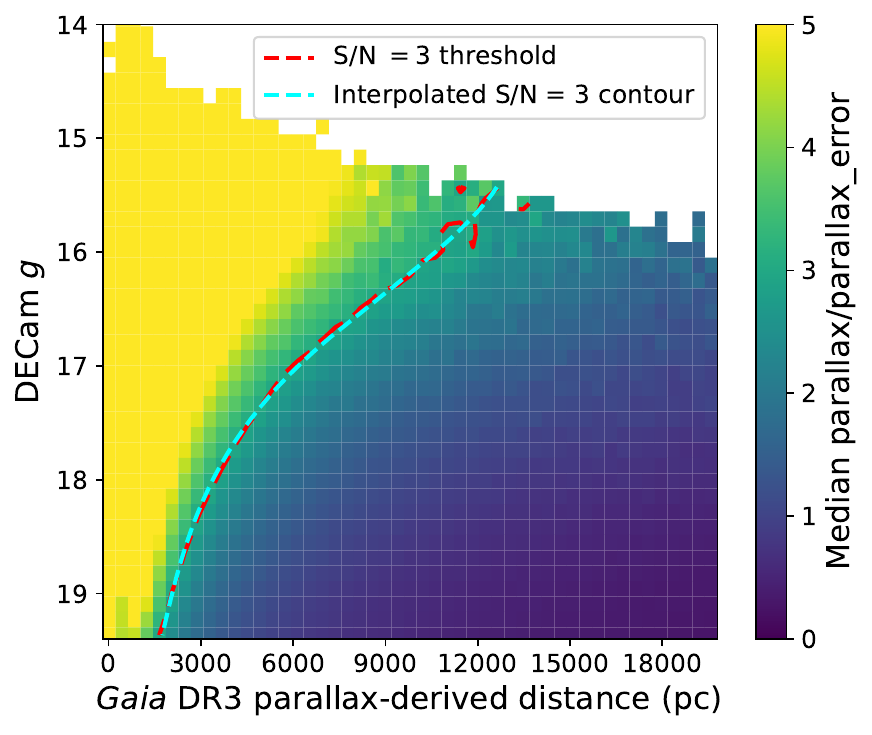}
    \caption{A 2D histogram of apparent DELVE DR2 $g$ magnitude vs. \textit{Gaia} DR3 parallax-derived distances for sources in our catalog, where each pixel is colored by the median significance of the parallax measurement (\texttt{parallax/parallax\_error}).
    The cyan dashed line indicates the interpolated contour at S/N = 3, tracing the typical distance out to which \textit{Gaia} DR3 yields $3\sigma$ parallax detections as a function of apparent magnitude. Stars to the left of this contour with unresolved parallaxes (i.e., whose distances should otherwise be detected) are likely distant RGB stars, enabling RGB/MS discrimination for a subset of distant stars discussed in Section~\ref{sec:derivationdistances}.}
    \label{fig:gaiasensitivity}
\end{figure}

In Figure~\ref{fig:gaiasensitivity}, we plot a 2D histogram of apparent DECam $g$ magnitude vs. \textit{Gaia} DR3 parallax for sources in our catalog, where each pixel is colored by the median significance of the parallax measurement (i.e., \texttt{parallax/parallax\_error}).
The cyan dashed line indicates the contour that traces the pixels at S/N = 3, showing the typical distance out to which \textit{Gaia} DR3 detects 3\,$\sigma$ parallaxes as a function of apparent magnitude $g$.
For stars with resolved parallaxes (i.e., \texttt{parallax/parallax\_error} $>$ 3), we convert the \textit{Gaia} DR3 parallax to a distance ($d_{\text{Gaia,kpc}}$ = 1/\texttt{parallax}) and compare this to the photometric distances inferred under each assumption. 
The star is classified as MS or RGB based on the photometric distance that is most consistent with the parallax-based distance. 

If the \textit{Gaia} DR3 parallax is unresolved (i.e., S/N $< 3$), we can still discriminate distant RGB stars from MS stars via two methods. 
First, if $d_{\text{MS}}$ is to the left of the curve in Figure~\ref{fig:gaiasensitivity} and the parallax is unresolved, then the star is classified as a distant RGB star since a parallax ought to have been detected if it was on the MS.
Second, if the inferred tangential velocity of the star based on the RGB distance and \textit{Gaia} DR3 proper motion is $> 600$\,km\,s$^{-1}$, then the star is classified as an MS star, due to this being larger than the escape velocity of the Milky Way, even in the solar neighborhood \citep{kh+21}.
In all other cases, the star is classified as having an ambiguous evolutionary state.

\subsection{Failure modes and contaminants}
\label{sec:failures}

Several classes of sources produce spurious photometric metallicities.
In this subsection, we describe some of these failure modes and the checks that we implement to identify potential contaminants.
We note that the first check against these failure modes is simply to require that the inferred photometric metallicities are within the bounds of our grid (i.e., [Fe/H]$_{\text{RGB}}$, [Fe/H]$_{\text{MS}} > -5.0$ and [Fe/H]$_{\text{RGB}}$, [Fe/H]$_{\text{MS}} < 1.0$), since contaminating sources will generally have colors beyond the metallicity contours spanned by MS and RGB stars.

\begin{figure}[htbp]
    \centering
    \includegraphics[width=1\columnwidth]{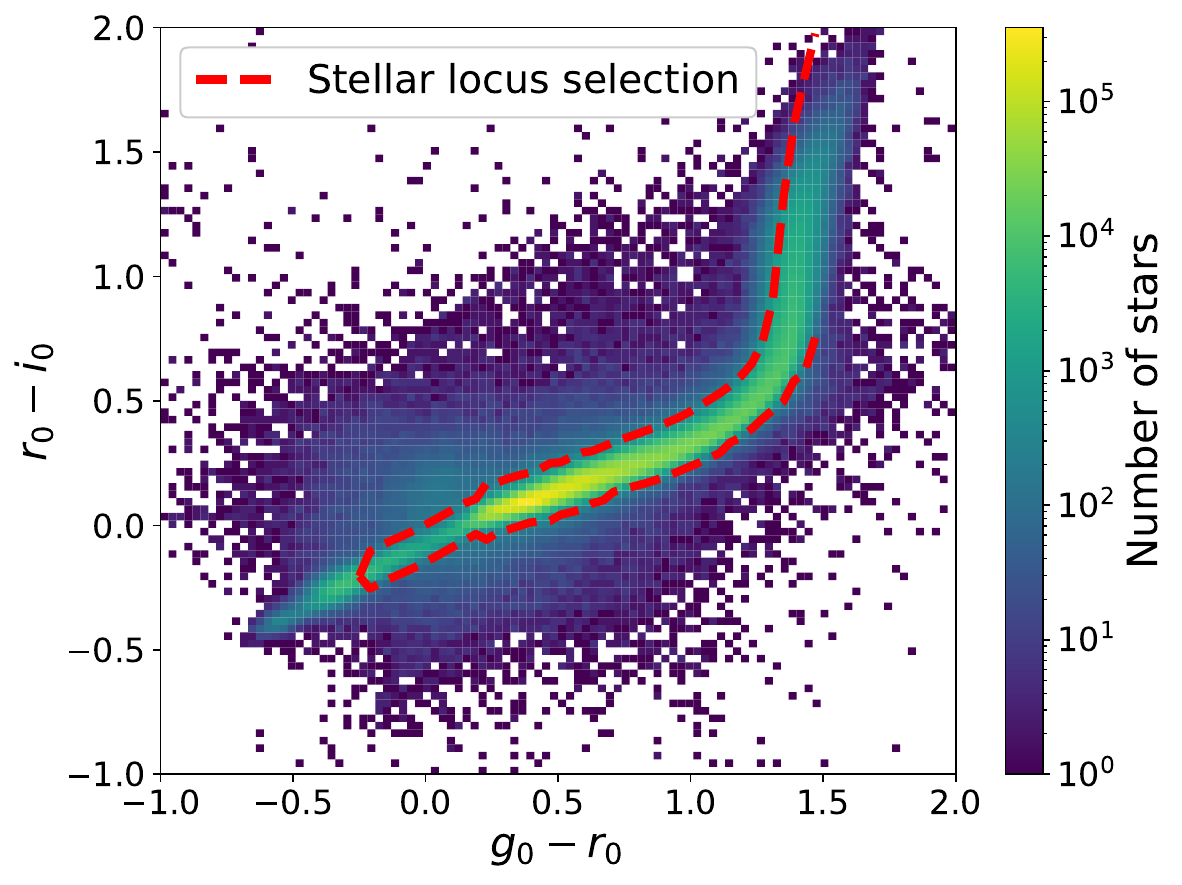}
    \caption{A 2D histogram of de-reddened broadband DELVE DR2 colors $(r_0 - i_0)$ vs. $(g_0 - r_0)$ for sources in the MAGIC catalog, colored by the number of stars per pixel. 
    The RGB and MS stars trace a tight locus (outlined in red), while contaminants (e.g., AGN, variable sources with discrepant multi-epoch broadband photometry) can scatter off this locus. 
    This locus in broadband color-color space corresponds to the selection described in Section~\ref{sec:failures} to remove interlopers with erroneous metallicities in our catalog.}
    \label{fig:broadbandcolorcolor}
\end{figure}

\textbf{Extragalactic contamination}-- Unresolved galaxies and active galactic nuclei (AGN) can contaminate our catalog and will have spurious metallicities, since the grid for inferring metallicities is derived from synthetic stellar photometry. 
We first exclude likely galaxies from having metallicities in the catalog by requiring \texttt{extended\_class\_g} $< 2$ in DELVE DR2 \citep{delvedr2} before any metallicity derivation.
This flag is a purely morphological selection from broadband DECam $g$-band images, and ought to reliably flag extended sources in the magnitude ranges for which we derive metallicities ($g \lesssim 21$). 
We supplement this with a broadband color-color selection in $(g - r)_0$ vs. $(r - i)_0$ space, where RGB and MS stars trace a tight relation and galaxies can occupy a broader locus from the stellar sequence (see Figure~\ref{fig:broadbandcolorcolor}).
The blue cutoff in this selection is motivated by the blue end of the horizontal branch in old, metal-poor Milky Way satellites \citep[][]{cmd+23}.

\textbf{Variable sources}-- Time-varying sources will have spurious photometric metallicities in the MAGIC catalog, as the CaHK imaging will almost certainly be sampled at discrepant phases from the broadband $g,i$ photometry. 
We flag known variable sources in our catalog via cross-match with the \textit{Gaia} DR3 variable star catalog \citep{gaiavar1, gaiavar2}.
Before any spectroscopic follow-up, we also cross-match candidate lists with the SIMBAD astronomical database \citep{simbad}, and exclude sources flagged as variable sources (e.g., eclipsing binaries, Cepheids, RR Lyrae, AGN, quasars).
Additionally, the broadband color-color selection can exclude variable sources since the broadband $g, r, i$ imaging in DELVE DR2 are typically obtained over multiple epochs. 
Any significant variability between these epochs causes stars to have inferred broadband colors that lie outside the locus in Figure~\ref{fig:broadbandcolorcolor}, providing an independent check against variable sources. 
We note that upcoming time-domain data from the Vera Rubin Observatory Legacy Survey of Space and Time (LSST; \citealt{lsst}) will likely minimize this failure mode. 

\begin{figure}[htbp]
    \centering
    \includegraphics[width=1\columnwidth]{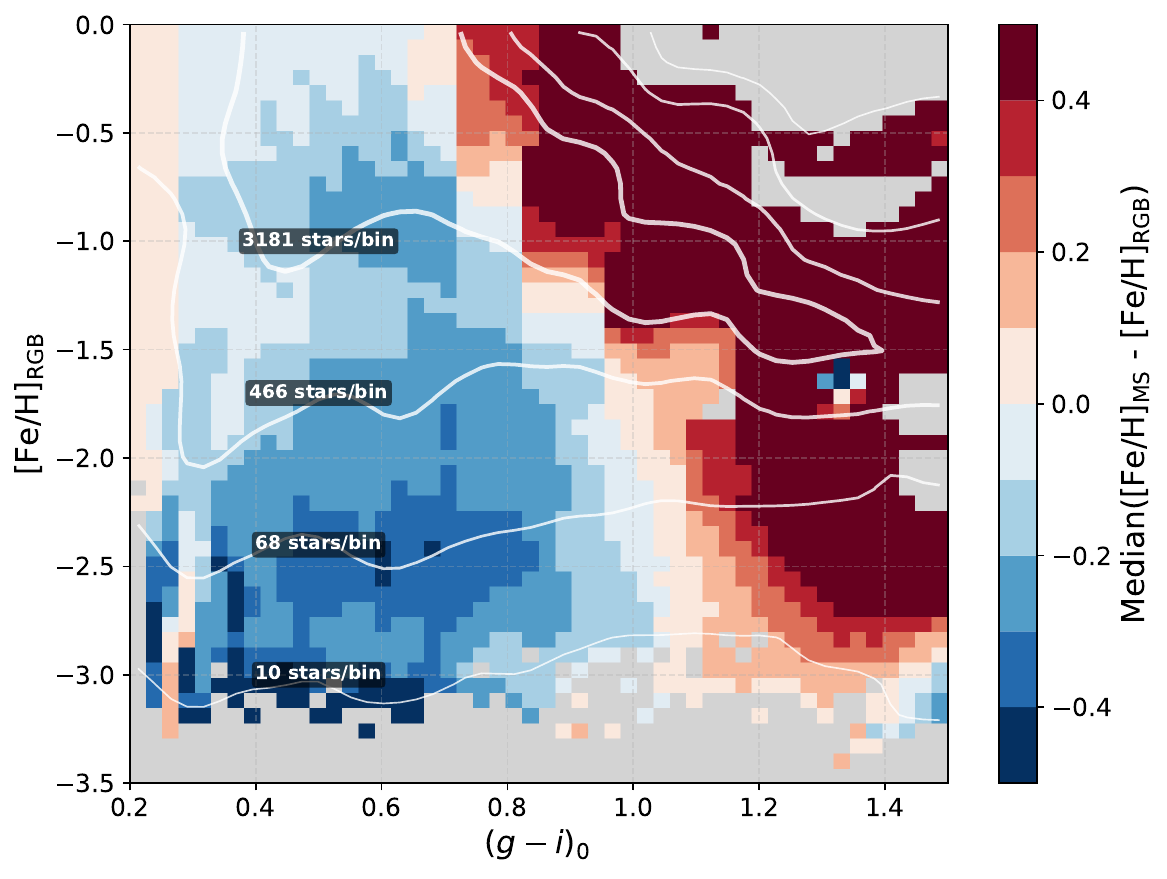}
    \caption{A 2D histogram visualizing the impact of dwarf/giant misclassification on inferred photometric metallicity in our methodology (Section~\ref{sec:derivationfeh}).
    Each pixel shows the median difference between the MS and RGB-assumed metallicities (i.e., [Fe/H]$_{\text{MS}}-$[Fe/H]$_{\text{RGB}}$) as a function of $g_0 - i_0$ and the RGB-assumed metallicity ([Fe/H]$_{\text{RGB}}$). 
    Contours are overlaid corresponding to the number of stars in each bin in the catalog of vetted metallicities described in the last paragraph of Section~\ref{sec:failures}.
    For cool stars ($(g-i)_0 \gtrsim 1.0$) and at low metallicities ([Fe/H] $\lesssim -3.0$), the discrepancy can be significant ($>0.5$\,dex).}
    \label{fig:dwarfgiantmismatch}
\end{figure}

\textbf{Dwarf/giant misclassification}-- As described in Section~\ref{sec:derivationdistances}, our classification of stars as being on the MS or RGB relies on \textit{Gaia} DR3 parallax constraints, which have limited utility when $g \gtrsim 19.5$. 
Figure~\ref{fig:dwarfgiantmismatch} illustrates the impact of dwarf/giant misclassification on the inferred metallicity as a function of color and metallicity. 
The misclassification of sources can significantly impact the metallicity estimate, especially for cool stars (e.g., $(g-i)_0 > 0.9$) and at low-metallicities. 
In the MAGIC catalog, we flag the adopted classification for each source (RGB, MS, Ambiguous), and propagate [Fe/H]$_{\text{RGB,MS}}$, $d_{\text{RGB,MS}}$ to the final [Fe/H] or $d$ if the \textit{Gaia} parallax permits a classification. 
Stars without a clear classification are reported with two metallicity and distance estimates, corresponding to an RGB or MS status.
We note for completeness that stars not in these evolutionary states (e.g., horizontal branch stars, blue stragglers) may contaminate a subset of the bluest stellar sources in the dataset.
The most extreme of these fail the broadband color-color flag (i.e., Figure~\ref{fig:broadbandcolorcolor}), and are likely to be flagged due to having unphysical photometric metallicities, as our grid of synthetic photometry is not optimized for these stars.

\textbf{High-reddening regions}-- Regions of high reddening are known to potentially lead to spurious photometric metallicity estimates. 
This is due to several reasons, including the presence of young stars with Ca\,{\sc ii}~H\&K emission features and the fact that the foreground line-of-sight reddening for each source is required for accurate de-reddening, as opposed to the treatment in Section~\ref{sec:processing}.
The net effect of the latter is to artificially decrease the photometric metallicity estimate of sources in high-reddening regions.
Consequently, we conservatively exclude regions of high reddening ($E(B-V)_{\text{SFD98}} > 0.2$ in \citealt{sfd+98}) in our analyses at this time, and prioritize low-metallicity sources in low reddening regions when performing spectroscopic follow-up. 
In addition, young or active stellar sources with emission features over the CaHK band will likely contaminate any photometric metallicities in our narrow-band photometry.
For this reason, we exclude the central 3$^\circ$ of the Small Magellanic Cloud (SMC)\footnote{The same would also be applied to the LMC, if MAGIC imaging extended to its central regions. DELVE DR2 photometry can be suspect in the center of the MCs, and this region will be flagged in future MAGIC data releases.} in our data where these populations are likely concentrated. 

\textbf{Photometric quality flags}-- In addition to the above, we propagate the standard Source Extractor quality flags (i.e., \texttt{FLAGS}, \texttt{IMAFLAGS\_ISO}) to the final catalog. 
Sources with \texttt{FLAGS} $> 3$, \texttt{IMAFLAGS\_ISO} $> 0$, or \texttt{SPREAD\_MODEL} $< -0.003$, indicating saturation, bad pixels, or cosmic ray contamination, are flagged and excluded from any photometric metallicity analyses.

In further analyses and comparisons, we require the following quality cuts, motivated by this section: (1) Consistency with the broadband color-color stellar locus (Figure~\ref{fig:broadbandcolorcolor}); (2) No extended sources (via DELVE DR2 $g$-band morphology; \citealt{delvedr2}); (3) No sources flagged as variable in \textit{Gaia} DR3 \citep{gaiavar1}; (4) Sources in relatively low reddening regions ($E(B-V) < 0.2$ in \citealt{sfd+98}); (5) No sources within 3$^\circ$ of the MCs.
In addition, we require the following further conditions to ensure high-quality metallicities: (1) metallicity uncertainties $\sigma_{\text{[Fe/H]}} < 0.5$\,dex; (2) an unambiguous RGB/MS classification based on Section~\ref{sec:derivationdistances}; (3) $0.2 < (g-i)_0 < 1.5$ to exclude the coolest stars and regions significantly bluer than the main-sequence turnoff; (4) [Fe/H]$_{\text{MAGIC}} > -4.0$ given the extreme rarity of genuine stellar sources at metallicities below this regime\footnote{Although, stars in this regime can still be spectroscopically targeted in a bespoke manner (see Section~\ref{sec:initialemp}).}.
In the current dataset, N=3,442,434 stars can be categorized as MS or RGB stars with valid metallicity values based on these criteria, which corresponds to around half of the stars that pass the aforementioned quality criteria when $g < 19.5$.

\begin{figure}[htbp]
    \centering
    \includegraphics[width=1\columnwidth]{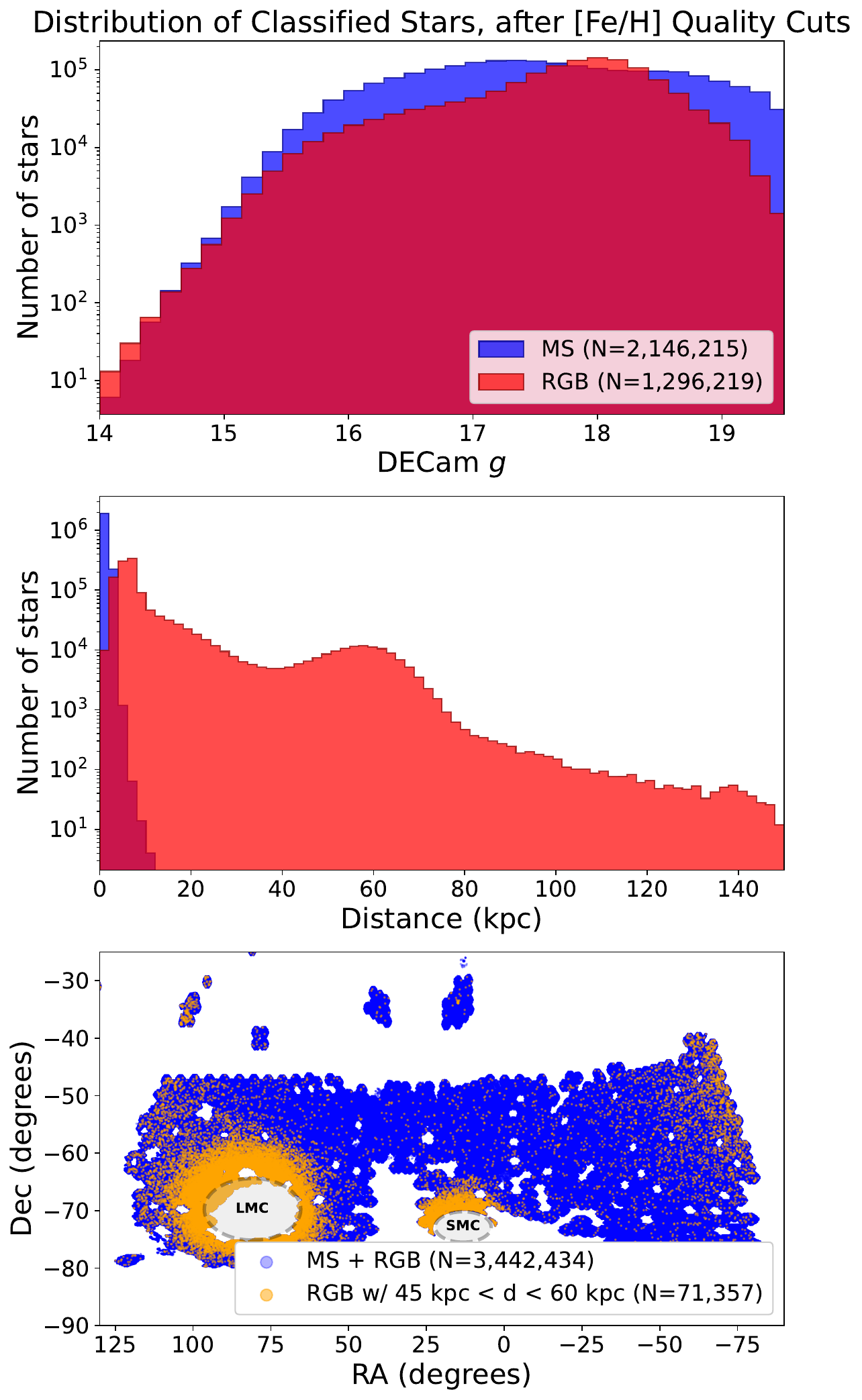}
    \caption{Diagnostics of sources in the MAGIC catalog classified as MS or RGB with valid photometric metallicities and passing the quality cuts described in Section~\ref{sec:failures} (e.g., broadband color-color selection, exclusion of variable sources in \textit{Gaia} DR3, photometric quality flags, and requiring metallicities within the bounds of the synthetic grid).
    Top: A histogram of MS and RGB stars as a function of apparent $g$ magnitude in DELVE DR2. 
    Middle: A histogram of MS and RGB stars as a function of inferred photometric distance from Section~\ref{sec:derivationdistances}.
    An apparent bump appears at the approximate distance of the MCs.
    Bottom: An on-sky distribution of all sources in the above panels (blue) in the vicinity of the MAGIC survey footprint, with stars classified as being RGB stars between 45\,kpc and 60\,kpc in orange.
    The MCs are clearly recovered, with slight contamination from likely misclassified MS stars when approaching the high source density near the Milky Way disk at the eastern and western ends of the footprint. }
    \label{fig:dist_spread}
\end{figure}

\begin{figure*}[htbp]
    \centering
    \includegraphics[width=\textwidth]{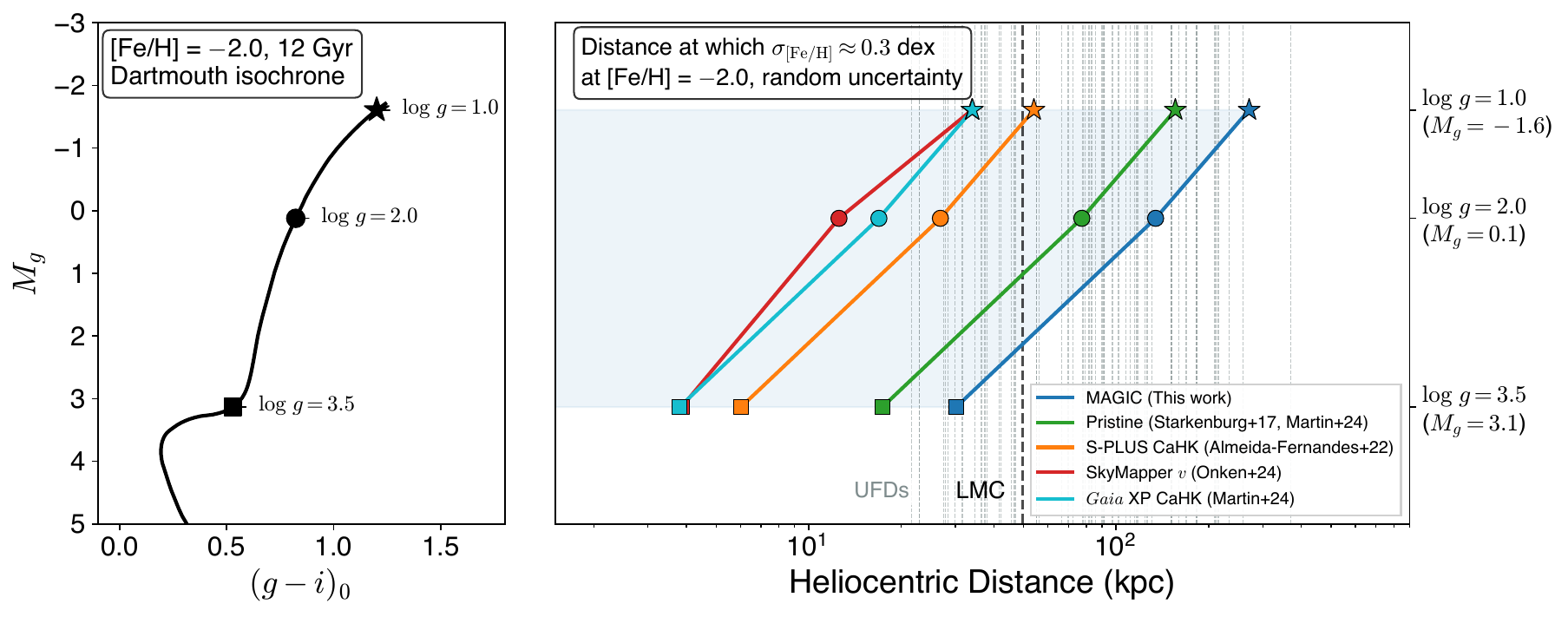}
    \caption{Left: A [Fe/H] = $-2.0$, 12\,Gyr Dartmouth isochrone \citep{dcj+08} with three locations on the RGB indicated with a star ($\log\,g = 1.0$), a circle ($\log\,g = 2.0$), and a square ($\log\,g = 3.5$). 
    Right: Heliocentric distance at which the random photometric metallicity uncertainty reaches $\sim 0.3$\,dex at [Fe/H]$=-2.0$ for each of the three reference RGB locations. 
    The MAGIC curve (blue) is computed using the metallicity uncertainties in the current catalog, and the Pristine (green; \citealt{smy+17}), S-PLUS (orange; \citealt{ash+22}), and synthetic CaHK from \textit{Gaia} XP spectra (purple; \citealt{msy+24}) are computed by scaling the MAGIC curve relative to the reported S/N of these surveys (see Section~\ref{sec:processing}). 
    The curve for the SkyMapper $v$ photometry (red; \citealt{owb+24}) is generated by re-deriving the grid of synthetic photometry using the SkyMapper filterset, and their reported $v$ depth.
    The vertical dashed lines indicate distances to known Milky Way UFDs and the LMC in the Local Volume Database \citep{pace+25}.
    Note that these curves only reflect the photometric uncertainty, and do not include systematic effects from dwarf/giant mis-classification. }
    \label{fig:sensitivity}
\end{figure*}

The magnitude distribution and inferred distance distribution of these stars are shown in Figure~\ref{fig:dist_spread}. 
The top panel demonstrates that our survey probes stars classified as RGB stars out to ${\sim}150$\,kpc. 
The middle panel shows a prominent overdensity of RGB stars at $\sim50$\,kpc that corresponds to the MCs.
The on-sky plot in the bottom panel of Figure~\ref{fig:dist_spread} shows that RGB stars at 45\,kpc to 60\,kpc clearly recover the outskirts of the MCs, with slight contamination from presumably misclassified MS stars due to the high source density of the Milky Way disk at the eastern and western edges of the footprint. 
Figure~\ref{fig:sensitivity} shows the distances at which stars at various points on the RGB have random metallicity uncertainty (i.e., ignoring dwarf/giant discrimination) of $\sim0.3$\,dex at [Fe/H]$_{\text{MAGIC,RGB}} = -2.0$ in the latest MAGIC catalog\footnote{v251003}. 
We also show this sensitivity curve scaled to the reported depths of other surveys (see Section~\ref{sec:processing}), and re-generated for the SkyMapper filter set for their metallicity-sensitive $v$ filter.
The increased depth of MAGIC meaningfully enhances the distance to which low metallicity structures can be probed with the catalog.
The spatial distribution of low metallicity stars in our catalog, including fainter sources, is further discussed in Section~\ref{sec:substructure}.

\subsection{External validation of MAGIC metallicities}
\label{sec:validation}

\begin{figure*}[htbp]
    \centering
    \includegraphics[width=1\textwidth]{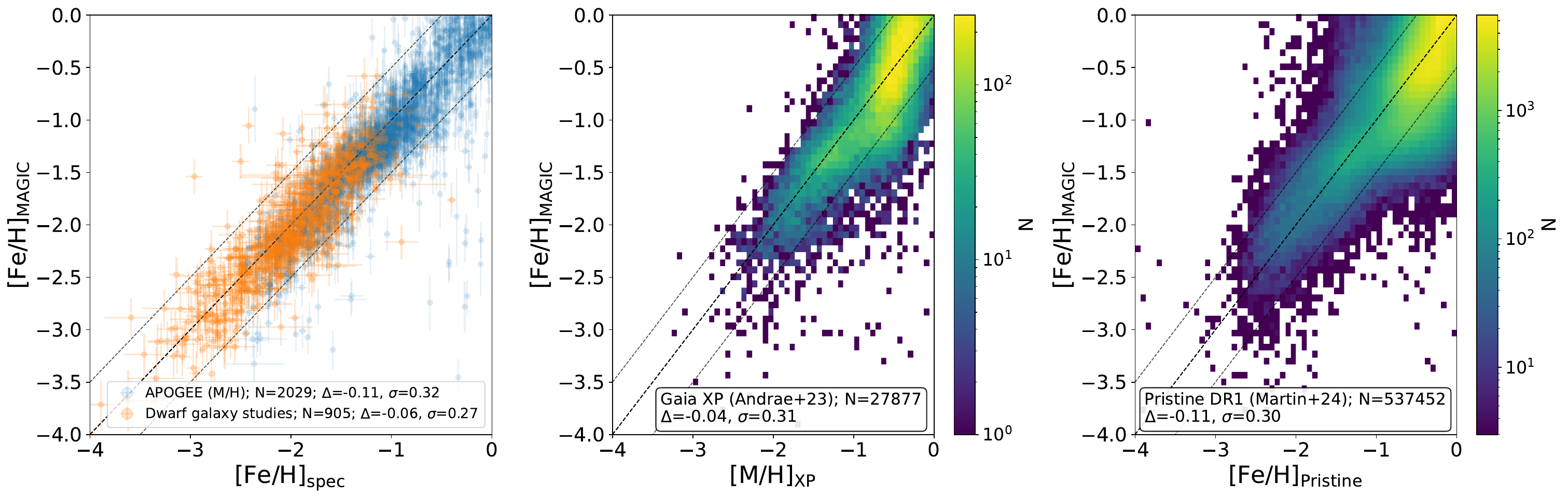}
    \caption{Comparison of MAGIC photometric metallicities to literature metallicities for sources classified as RGB or MS stars in the current catalog (see Section~\ref{sec:derivationdistances}) and passing the quality cuts described in Section~\ref{sec:failures}. 
    Left: A metallicity comparison with respect to [M/H] in APOGEE DR17 \citep{aaa+22} (blue points), and literature dwarf galaxy metallicities for selected systems to gauge performance in the low-metallicity regime \citep{sdl+15, koposov+15, li+2018carina, cfj+18, lsk+18, fcb+19, sle+20, csf+22, chiti2023, tsa+23, cml+24, oh+24, ljs+24, hlp+24, plj+25, tba+25, gpp+26}. See Appendix~\ref{app:dwarfxmatch} for individual sample sizes.
    Middle and right: Metallicity comparisons with respect to metallicities from \textit{Gaia} XP spectra from \citet{andrae2023}, and Pristine DR1 \citep{msy+24}, respectively. The middle and right panels are shown as 2D histograms colored by the number of stars per pixel. 
    The dashed line in each panel indicates the one-to-one line to guide the eye, with the grey lines denoting offsets of $\pm0.5$\,dex.
    We include a detailed discussion of these comparisons in Section~\ref{sec:validation}.}
    \label{fig:validation}
\end{figure*}

In this section, we compare the metallicity measurements in our catalog to literature values to assess their performance. 
We note that the metallicity performance in targeted regimes has already been demonstrated in early MAGIC publications \citep[see Section~\ref{sec:lit};][]{bcl+25, plc+25, cpp+25, apk+26}, including explicit comparisons to spectroscopic metallicities for the Sculptor dSph in \citet{bcl+25} and the Crater II dSph in \citet{apk+26}. 
Here, we present metallicity comparisons using the MAGIC data collected so far to metallicities in APOGEE DR17 \citep{SDSSIII-2011, SDSSIV-2017, msf+17, aaa+22} and a number of studies presenting spectroscopic metallicities in dwarf galaxies (Figure~\ref{fig:validation} caption), the \textit{Gaia} XP spectrum-derived metallicities in \citet{andrae2023}, and metallicities from Pristine DR1 \citep{msy+24} which are derived from synthetic narrow-band CaHK photometry using the \textit{Gaia} XP data \citep{gaiaxp}. 
We note that APOGEE in particular derives metallicities using infrared spectroscopy, distinct from the Ca\,{\sc{ii}}~H\&K lines used in MAGIC.

Figure~\ref{fig:validation} shows a three-panel comparison of the metallicities in our catalog versus literature measurements. 
The panels show comparisons to APOGEE DR17 spectroscopic metallicities (left; \citealt{aaa+22}), \textit{Gaia} XP photometric metallicities (middle: \citealt{andrae2023}), and Pristine DR1 (right; \citealt{msy+24}) for the full sample passing basic quality cuts (e.g., broadband color-color cut, no variable stars; see Section~\ref{sec:failures}), for all sources that can be classified as RGB or MS stars.
The comparison to APOGEE DR17 yields a low median offset of $-0.11$\,dex with a scatter of $0.32$\,dex, somewhat driven by a handful of outliers for which we measure notably lower metallicities.
Overlaid on this plot are metallicities from studies of various dwarf galaxies with classified MS/RGB stars in the current catalog \citep[see Appendix~\ref{app:dwarfxmatch};][]{sdl+15, koposov+15, li+2018carina, cfj+18, lsk+18, fcb+19, sle+20, csf+22, chiti2023, tsa+23, cml+24, oh+24, ljs+24, hlp+24, plj+25, tba+25, gpp+26}.
While the individual sample sizes tend to be small for the most metal-poor dwarf galaxies, they provide a cross-check on the metallicity performance in the most metal-poor regime ([Fe/H] $\lesssim -2.0$), and on average show reasonable scatter (${\sim}0.27\,$dex, with $\sigma=0.25\,$dex when $-4.0~<$~[Fe/H]$_{\text{MAGIC}}$~$<~-2.0$).
These additional datasets are important to spectroscopically cross-verify metallicity performance, as APOGEE DR17 only derives metallicities down to $\text{[Fe/H]}\approx-2.5$.

We note a slight systematic at the metal-rich end of the comparison to Gaia XP-based metallicities in \citet{andrae2023}. As MAGIC has largely focused on targeting and characterizing the metal-poor Milky Way halo, metallicity systematics at the metal-rich end have not been prohibitive for ongoing work, but are a source of investigation to be addressed in future data releases. 
For instance, the inclusion of $\log\,g$ values from the [Fe/H]~=~$-1.0$ isochrone (see Section~\ref{sec:derivationfeh}) alleviates this systematic relative to \citet{andrae2023}, but introduces a systematic relative to APOGEE.
Overall, the comparisons to the metallicity catalogs in \citet{andrae2023} and Pristine DR1 \citep{msy+24} show reasonable offsets ($\lesssim 0.11$\,dex) and scatter $\sigma \approx 0.3$\,dex.
Metallicity comparisons for targeted studies using the MAGIC dataset (e.g., \citealt{bcl+25, apk+26}) typically show lower scatter (${\sim}0.2$\,dex) as they focus on clean samples of metal-poor RGB stars, which reduces the likelihood of catastrophic outliers. 
MAGIC photometric distances will be evaluated and presented in an upcoming publication of a metallicity map of the Milky Way, but are implicitly tested in the substructure maps presented in Section~\ref{sec:substructure}.

\section{Early Science Results}
\label{sec:science}

In this section, we overview results from the MAGIC survey that are already in the literature \citep{bcl+25, plc+25, cpp+25, apk+26, dcf+26}, demonstrating the feasibility of our scientific aims (Section~\ref{sec:lit}).
In addition, we present several initial results that touch on future prospects (Section~\ref{sec:retii}--~\ref{sec:initialemp}).

\subsection{Summary of public results}
\label{sec:lit}

At the time of this writing, a total of seven publications have been submitted or published using data collected through MAGIC or MAGIC-affiliated P.I. programs processed with the MAGIC pipeline (see Section~\ref{sec:strategy}). 
Five of these publications used these data in a primary capacity, and we summarize those works below. 
The two additional publications \citep{plj+25,tcp+25} used MAGIC data in a limited capacity to cross-verify the membership of stars in UFDs.

\cite{bcl+25} presented the largest sample of members for the Sculptor dSph with measured metallicities ($N\approx3800$). 
This sample allowed an external cross-validation of MAGIC photometric metallicity estimates with the ${\sim}1000$ available spectroscopic metallicities \citep{tsa+23} within the system. 
Our sample improved the determination of the MDF and metallicity gradient of Sculptor, due to the depth and spatially unbiased coverage of the survey around the dwarf galaxy.
We measured a broken metallicity profile, with a steep gradient within ${\sim} 1 \ r_h$ ($\sim11.17\arcmin$; \citealt{mcs+18}) of $-3.26 \pm 0.18 \rm \ dex\,deg^{-1}$, and $-0.55 \pm 0.26 \rm \ dex\,deg^{-1}$ beyond the break radius. 
Moreover, we detected six new extremely metal-poor candidates, demonstrating the utility of the MAGIC survey to target the most metal-poor stars in Milky Way satellites for future programs. 

\citet{plc+25} presented high-resolution ($R>20,000$) spectroscopic observations for six metal-poor stars with [Fe/H]~$<-3.0$ (including one carbon-enhanced star with [Fe/H]~$=-4.12$), selected from MAGIC. 
We confirmed the accuracy of the photometric metallicities (median $\Delta{\rm{[Fe/H]}}=-0.05$) and determined chemical abundances for up to 16 elements per star, from carbon to barium.
A kinematic/dynamical analysis indicated that all program stars belong to the distant Milky Way halo population ($\gtrsim30$\,kpc), including three stars with high-energy orbits that might have been associated with the MCs, one consistent with being a member of the Sagittarius stream \citep{igi+94, msw+03}, and another two showing kinematics consistent with the Gaia-Sausage/Enceladus dwarf galaxy merger.
Our detection of a new [Fe/H] $< -4.0$ star at a large distance ($\approx35$\,kpc) in the Milky Way halo demonstrated the capacity of the survey to recover the lowest metallicity stars in new regimes. 

\citet{cpp+25} presented the discovery and initial chemical characterization of the star with the lowest iron abundance (a non-detection at [Fe/H] $< -4.63$) known in any UFD.
The star, designated PicII-503, is located in the outskirts of the Pictor II UFD and was targeted through its extremely low metallicity in the MAGIC catalog ([Fe/H]$_{\text{RGB}} \approx -3.8$). 
Notably, the only two elements with detected abundances in the star based on $R\approx10{,}000$ VLT/X-Shooter \citep{vdd+11} spectroscopy were calcium ([Ca/H]~=~$-5.2$~+/$-$~0.31/0.5) and carbon ([C/H] = $-1.44 \pm 0.38$). 
The extreme paucity of heavy elements (e.g., [Fe/H] $< -4.63$, [Ca/H] $ = -5.2$, [Mg/H] $< -4.42$) in PicII-503 shows it to clearly preserve enrichment by Population III supernovae within a UFD, demonstrating that extreme carbon enhancement is a pathway by which the first stars can enrich primordial, small-scale structures. 
We note that the Pictor~II UFD is a satellite of the LMC \citep{plj+25}.

\citet{apk+26} used MAGIC metallicities to identify RGB stars in the Crater II dSph's stellar stream. 
The paper presents DECam CaHK photometry of 162 candidate stars across 128 ${\rm deg}^2$, measuring the first significant metallicity gradient of $-0.34 \pm 0.17 ~{\rm dex}~{\rm deg}^{-1}$ from 124 candidates in the main body and validating $N$-body tidal disruption model predictions for a near and far aligned stream component using 37 candidates in the tidal tails. 
The star counts indicate a scaled surface brightness of ${\sim} 36 ~{\rm mag}~{\rm arcsec}^{-2}$ in the stream and a stream width of approximately $0.8^\circ$ (${\sim} 1.63$ kpc) that is consistent with both cored and cuspy dark matter halo models.
This advances studies of Crater II's unusual properties, showing how the MAGIC survey will enable unprecedented sensitivity to faint, diffuse tidal features in metal-poor systems.
This work also cross-validated MAGIC metallicities with those in spectroscopic studies of Crater II \citep{jkl+21,ljl+25}.

\citet{dcf+26} investigated the morphology of the Jet stellar stream, using DECam MAGIC CaHK photometric metallicities in tandem with \textit{Gaia} DR3 \citep{gaiadr3} photometry to identify candidate stream members. Of the 611,961 observed stars in the initial catalog, 213 candidate members were identified, and the purity of the candidate selection was evaluated in comparison to a sample of spectroscopically confirmed Jet stream members from the Southern Sky Stellar Stream Spectroscopic Survey (\citealt{lkz+19}). The inclusion of MAGIC metallicities reduced non-member contamination from $\sim76\%$ to $\sim18\%$, producing a significantly cleaner Jet Stream member sample than could be achieved with CMD and proper-motion cuts alone. This enabled the morphological characterization of the Jet stream, corroborating a stream width of $\sim0.18^{\circ}$ as suggested by \citet{fsd+22}, and revealed a fanning of the stream toward the end of the stream farther away from the Milky Way bar. This demonstrates the applicability of the MAGIC survey in identifying members of distant, low-surface brightness systems in the metal-poor regime, providing important infrastructure for further spectroscopic follow up.

\subsection{An r-process enhanced star in the outskirts of Reticulum II}
\label{sec:retii}

Reticulum~II is a UFD that is likely bound to the LMC \citep[e.g.,][]{pel+22}, and hosts stars that show a dramatic enhancement in elements produced by the rapid neutron-capture process (i.e., $r$-process; \citealt{jfc+16, rmb+16, jsr+23}). 
The $r$-process enhancement in stars in Reticulum~II matches the highest level known among stars in the Milky Way halo ([Eu/Fe] $> +1.0$), known as $r$-II stars \citep[e.g.,][]{bc+05, hhb+20}. 
Consequently, Reticulum~II preserves enrichment from a single prolific r-process production event from the early universe (\citealt{jfc+16, bjd+21}).
This peculiar signature of Reticulum II stars principally allows one to tag likely members of Reticulum II in its far outskirts, since this level of $r$-process enhancement is uncommon among stars in the Milky Way halo (8\,\% of stars with [Fe/H] $< -2.0$ in the halo are $r$-II stars; \citealt{erf+20, hhb+20}).

\begin{figure*}[htbp]
    \centering
    \includegraphics[width=1\textwidth]{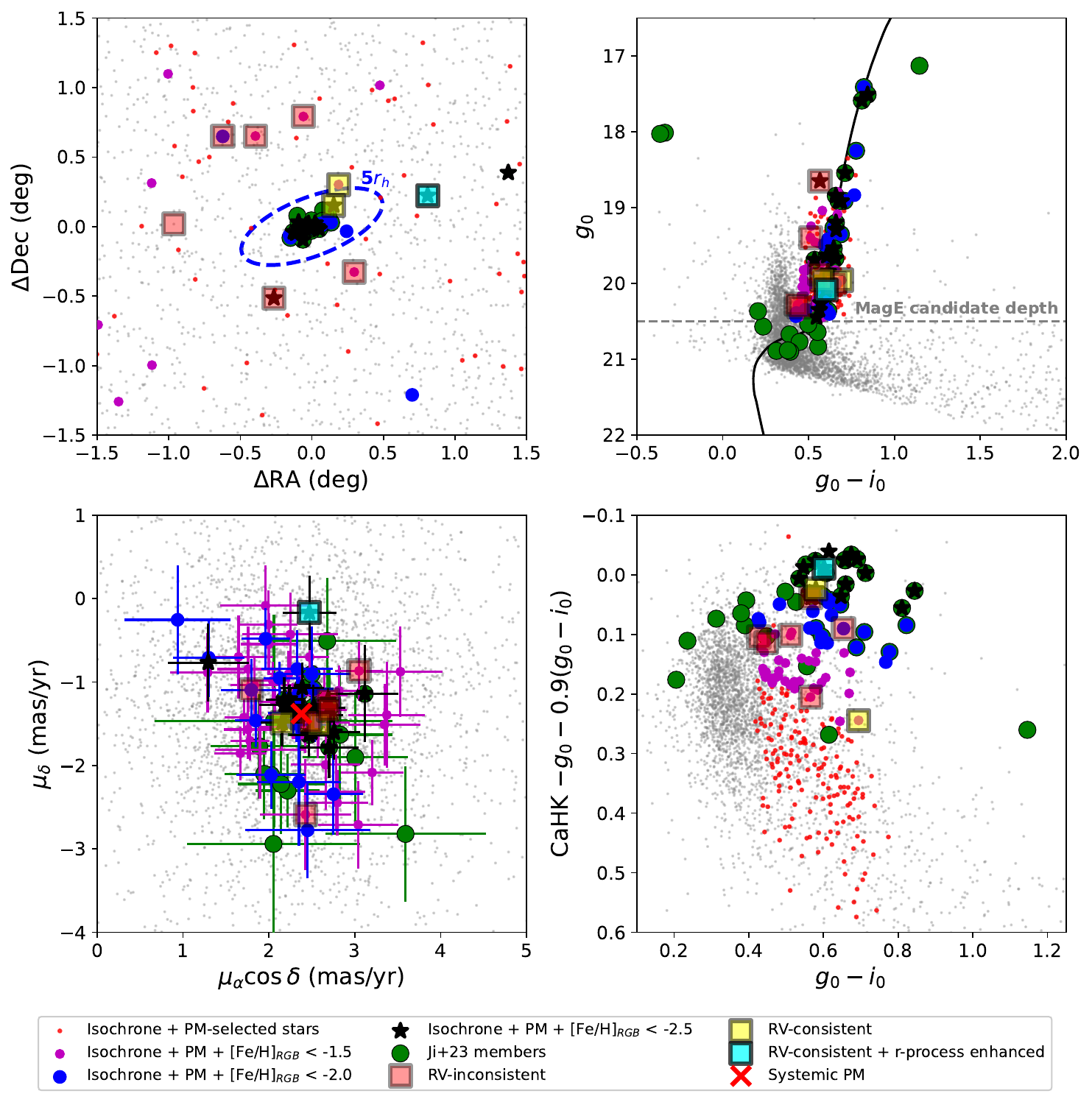}
    \caption{The selection of distant Reticulum II UFD candidate members from MAGIC, including a new r-process enhanced member at large radius. Top left: Spatial distribution of stars around the Reticulum II UFD, with the blue dashed ellipse denoting $5\,r_h$. Grey points show all sources in the MAGIC catalog, red points are stars consistent with a 12 Gyr, [Fe/H] =$-$2.5 Dartmouth isochrone \citep{dcj+08} at the distance of Reticulum II ($m-M=17.5$; \citealt{msc+18}) and the systemic proper motion of the system ($\mu_{\alpha}\cos{\delta} = 2.377 \pm 0.24$\,mas\,yr$^{-1}$, $\mu_{\delta} = -1.379 \pm 0.25$\,mas\,yr$^{-1}$; \citealt{pel+22}). 
    Larger magenta and blue circles, and black stars indicate the subsets with MAGIC [Fe/H]$_{\text{RGB}} < -1.5, < -2.0, \text{and} <-2.5$, respectively. Green circles are spectroscopically confirmed members in \citet{Ji+2023}. Square markers indicate the MagE follow-up outcomes. 
    Top right: Same markers and colors, but on a CMD using DELVE DR2 $g, i$ photometry. 
    The solid curve is the aforementioned isochrone at the distance of Reticulum II, and the horizontal grey dashed line at $g_0\approx20.5$ indicates the approximate depth of our MagE targeting. 
    Bottom left: Same markers, but on a plot of \textit{Gaia} DR3 proper motions \citep{gaiadr3}. 
    The red cross marks the systemic proper motion of Reticulum II. 
    Bottom right: Same markers, but on a color-color plot that incorporates the metallicity-sensitive CaHK filter on the y-axis. 
    The location of stars on this plot maps onto their metallicities (see Figure~\ref{fig:synthfeh}), illustrating the separation of stars belonging to the very metal-poor Reticulum II UFD from the more metal-rich foreground/background Milky Way stars.}
    \label{fig:retii}
\end{figure*}

As discussed in Section~\ref{sec:aims}, an open question is whether UFDs host extended populations of stars that may indicate their ongoing tidal disruption, or trace their underlying dark matter halos. 
We performed a search for stars in the outskirts of Reticulum II using MAGIC photometry (illustrated in Figure~\ref{fig:retii}), following standard search sequences for metal-poor stars in the outskirts of UFDs \citep[e.g.,][]{cfs+21, ljb+23, ocs+24}.
First, we selected stars with broadband DELVE DR2 photometry consistent with a 12\,Gyr, [Fe/H] $=-2.5$ Dartmouth isochrone \citep{dcj+08} placed at the distance modulus of Reticulum~II ($m-M = 17.5$; \citealt{msc+18}). 
Then, we selected stars with \textit{Gaia} DR3 proper motions consistent with that of the systemic motion of the system ($\mu_{\alpha}\cos{\delta} = 2.377 \pm 0.24$\,mas\,yr$^{-1}$, $\mu_{\delta} = -1.379 \pm 0.25$\,mas\,yr$^{-1}$; \citealt{pel+22}) at the 3\,$\sigma$ level, based on the proper motion uncertainties on individual stars. 
Finally, we selected stars that passed various metallicity thresholds in the MAGIC catalog, assuming RGB values (see top left of Figure~\ref{fig:retii}).
Only three stars appear in the outskirts ($>5\,r_h$) of Reticulum~II out to ${\sim}2^{\circ}$ that pass the above selection with MAGIC [Fe/H]$_{\text{RGB}} < -2.5$.
This cut represents the most stringent (i.e., purest) threshold for candidate selection; relaxing to [Fe/H]$_{\text{RGB}} < -2.0$ and [Fe/H]$_{\text{RGB}} < -1.5$ yield additional candidates (Figure~\ref{fig:retii}).

We obtained spectroscopy of nine stars in the outskirts of Reticulum~II at a range of metallicities with \textit{Gaia} DR3 proper motions consistent with membership using the MagE spectrograph \citep{mbt+08} on the 6.5\,m Magellan/Baade Telescope at Las Campanas Observatory on 2023 December 6--8, using the 0\farcs7 slit (yielding $R\approx6000$), with two candidates observed on 2023 October 14. 
We derived velocities for these stars by cross-correlating the observed stars with a template MagE spectrum of HD122563, exactly following the procedure in \citet{cpp+25}.
All but three of the observed candidates are excluded as candidate members based on their heliocentric radial velocities, given the velocities of Reticulum~II's member stars ($\approx63$\,km\,s$^{-1}$; \citealt{sdl+15, wmo+15}).
We note that most of our MagE targets were observed for completeness in the near-vicinity of Reticulum~II, and had metallicities near the metal-rich end of our candidate selection (i.e., $-2.5 < $[Fe/H]$_{\text{RGB}} \lesssim -1.5$). 
Initial targeting was also based on preliminary metallicities from the November 2023 version of the MAGIC catalog, which did not have the full suite of quality-checks (e.g., stellar locus selection) discussed in Section~\ref{sec:failures}.
In Table~\ref{tab:mageobservations}, we list the coordinates, brightnesses, MagE exposure times, and derived heliocentric radial velocities for stars in Figure~\ref{fig:retii}.

We derive spectroscopic metallicities for stars in Table~\ref{tab:mageobservations} using the strength of the Ca\,{\sc ii}~K line, following the metallicity calibration described in \citet{brn+99}.
This calibration involves mapping a measure of the strength of the Ca\,{\sc ii}~K line known as the KP index, along with the $B-V$ color, to a metallicity. 
Briefly, the KP index measures the pseudo-equivalent width of the Ca\,{\sc ii}~K line by directly integrating the flux across the feature, using a linear interpolation through continuum bands to estimate the background flux. 
Our implementation of this method exactly follows \citet{csf+18}.
The $B-V$ color was derived by matching DELVE DR2 $(g-i)_0$ colors to theoretically predicted $B-V$ colors using 12\,Gyr, [Fe/H]~=~$-1.5, -2.0, -2.5$ Dartmouth isochrones, based on the MAGIC photometric metallicity of the star \citep{dcj+08}.
As shown in Table~\ref{tab:mageobservations}, the resulting KP index-based metallicities are consistent with photometric metallicities in the MAGIC catalog, and three stars (RetII-406,  RetII-409, RetII-502) have radial velocities $3\sigma$ consistent with membership to Reticulum~II ($v_{\text{sys}} = 62.8\pm0.5$\,km\,s$^{-1}$, with dispersion $\sigma=3.3\pm0.7\,$km\,s$^{-1}$; \citealt{sdl+15}). 
{RetII-409} and {RetII-502} also show spectroscopic metallicities in the very metal-poor regime, consistent with being members of Reticulum II ($\langle$[Fe/H]$\rangle=-2.59\pm0.05$, with dispersion $\sigma=0.40\pm0.05$; \citealt{ljc+25}).

\begin{figure}[htbp]
    \centering
    \includegraphics[width=1\columnwidth]{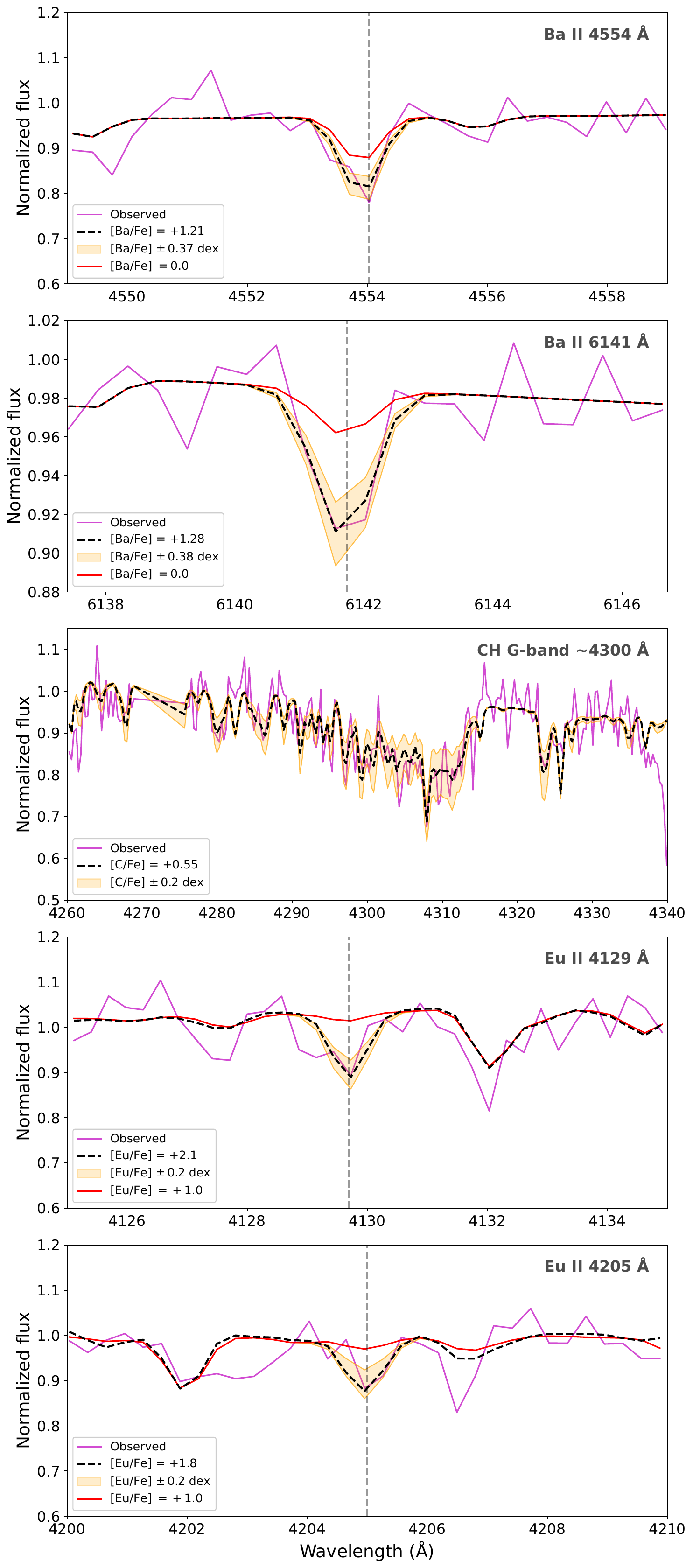}
    \caption{Spectral syntheses for RetII-409. Top: The observed Ba 4554\,{\AA} line (magenta), along with a synthetic spectrum at [Ba/Fe] = +1.21 (dashed line), synthetic spectra at [Eu/Fe] $\pm 0.37$\,dex (shaded orange), and threshold [Ba/Fe] = $+0.0$ (red). 
    The 0.37\,dex corresponds to the random uncertainty on the Ba abundance from this absorption feature, and [Ba/Fe] = 0.0 is the abundance threshold below which most Milky Way halo stars at the metallicity of RetII-409 reside (Figure~3 in \citealt{jfs+16}).
    Subsequent panels: Same as top, but for the Ba 6142\,{\AA} line, CH G-band, the Eu 4129\,{\AA} line, and the Eu 4205\,{\AA}.
    In the Eu panels, a synthetic spectrum at the $r$-II threshold of [Eu/Fe] = $+1.0$ is shown in red. 
    Given the continuum noise around the Eu features, we report a detection with an uncertain abundance in Table~\ref{tab:retiiabundances}.}
    \label{fig:retii409}
\end{figure}

Notably, the spectrum of RetII-409 displays evidence for europium (Eu) and prominent barium (Ba) absorption features, which are tracer elements for the $r$- and $s$-process.
Accordingly, we derive abundances for these elements in this star using standard 1D spectral synthesis techniques assuming local thermodynamic equilibrium (LTE) as described in \citet{cpp+25} using the Eu II 4129\,{\AA}, Eu II 4205\,{\AA}, Ba II 4554\,{\AA}, and Ba II 6141\,{\AA} lines. 
We use the MOOG spectral synthesis code \citep{s+73,sks+11}, ATLAS9 model atmospheres \citep{ck+03,k+05}, and the linelist from the \texttt{linemake} software \citep{psr+21} with entries from various sources \citep{mpv+14, slr+14, rdl+14, drl+14, bpr+17,phn+17,nist,dls+21}, within the Spectroscopy Made Harder (SMHR) software package \citep{c+14}.
Synthetic spectra were generated assuming the stellar parameters from the isochrone-matching described in the previous paragraph ($T_{\text{eff}} = 5476$\,K, $\log\,g = 3.46$), with a microturbulence of 1.5\,km\,s$^{-1}$ based on stars with similar $\log\,g$ \citep{fcj+13}.
Systematic abundance uncertainties were derived by propagating fiducial uncertainties of ($T_{\text{eff}} \pm 150$\,K, $\log\,g \pm 0.3$\,dex, $v_{\text{mic}} \pm 0.3$\,km\,s$^{-1}$). 
We show syntheses of the 4129\,{\AA}, 4205\,{\AA}, 4554\,{\AA}, and 6141\,{\AA} lines in Figure~\ref{fig:retii409}, in addition to the carbon G band at ${\sim}4300$\,{\AA}.

We present a detection of a high [Ba/Fe] = $1.24 \pm 0.28$, which clearly places RetII-409 in the regime of Reticulum II members versus Milky Way halo stars at the same metallicity ([Ba/Fe] $\approx 1.0$ vs. [Ba/Fe] $\lesssim 0.0$; see Figure~3 in \citealt{jfs+16}).
We also present a tentative detection of the $r$-process tracer element europium at the [Eu/Fe] $\approx 1.95$ level, independently suggesting that RetII-409 is an r-II star (i.e., with [Eu/Fe] $> +1.0$; see Figure~\ref{fig:retii409}). 
The comparatively less elevated [C/Fe]=$0.55\pm0.28$ shows that the elevated barium in this star is likely not from $s$-process production from a binary companion \citep[e.g.,][]{ran+05}, and rather originates from an $r$-process signature. 
This $r$-process enhancement provides chemical evidence that RetII-409 is associated with Reticulum II, and shows that the galaxy hosts a spatially extended population. 
RetII-409 lies at {$\sim8\,r_h$} and is along the elongation of Reticulum II, which aligns with its predicted tidal debris track if the UFD were losing mass due to tidal forces from the Milky Way accounting for the LMC (D. Erkal et al., priv. comm.).
The same debris track also predicts a several km\,s$^{-1}$ drift upward in the systemic radial velocity of Reticulum II members in the location of RetII-409, consistent with the observed higher radial velocity than the central members.  
We identify another candidate with MAGIC [Fe/H]$_{\text{RGB}} < -2.5$ along this same elongation (see Figure~\ref{fig:retii}), but the star was marginally too faint ($g=$20.49) for rapid spectroscopy with MagE. 
The adopted stellar parameters and element abundances of RetII-409 are shown in Table~\ref{tab:retiiabundances}.
Of the other velocity-consistent candidates, RetII-406 is too metal-rich ([Fe/H] = $-1.62\pm0.36$) for its [Eu/H] to act as a discriminant relative to Milky Way halo stars (see Figure~2 in \citealt{jfc+16}), and RetII-502 has a spectrum with insufficient S/N for chemical abundance analysis.
Accordingly, we list their membership status as just RV-consistent.

A demonstration that Reticulum II is influenced by Milky Way tidal forces is not necessarily surprising, given that it is likely at orbital pericenter \citep{lhb+21, hwp+21, pel+22}. 
However, as a likely satellite of the LMC, Reticulum II has not been subject to the tidal field of the Milky Way for long and its pericenter is sufficiently far from the center of the Milky Way (${\sim}37$\,kpc) that it has been argued to have not lost significant mass from tidal stripping from our Galaxy \citep[e.g.,][]{pel+22}.
Our observation establishes that photometric metallicities can be used to uncover stars in the faint outskirts of Reticulum II (${\sim}1$ RGB star from $r\approx0.5^{\circ}$ to $r\approx1^{\circ}$) to test these scenarios via targeted spectroscopy.

The additional detection of stars in the outskirts of Reticulum II ought to be able to disentangle the mechanism by which its faint outskirts formed.
Since most UFDs do not show $r$-process enhancement \citep{jsf+19}, theories proposing the formation of UFD stellar halos from merging UFDs would predict that the stars in the outskirts of Reticulum II ought not be $r$-process enhanced.
Alternatively, if stars in the outskirts of Reticulum II are $r$-process enhanced, this would indicate in-situ formation of stars and subsequent displacement to the outskirts via other mechanisms (e.g., tidal disruption, binary interactions).
The identification and spectroscopic characterization of larger samples of stars in the outskirts of Reticulum II are needed to test these hypotheses.
Such samples will require deeper proper motions than \textit{Gaia} DR3 for candidate selection, and would also be facilitated by deeper metallicity-sensitive photometry of the region.

\begin{deluxetable*}{lccccccccc}
\tablecaption{MagE observations of candidate Reticulum II stars in the outskirts of the system. \label{tab:mageobservations}}
\tablehead{
\colhead{Star ID} & \colhead{R.A.} & \colhead{Decl.} & \colhead{$g_0$} & \colhead{[Fe/H]$_{\text{MAGIC,RGB}}$} & \colhead{[Fe/H]$_{\text{MagE}}$} & \colhead{$t_{\text{exp}}$} & \colhead{S/N at $\sim$6500\,{\AA}} & \colhead{$v_{\text{hel}}$} & \colhead{Membership}\\
\colhead{} & \colhead{(deg)} & \colhead{(deg)} & \colhead{(mag)} & \colhead{(dex)} & \colhead{(dex)} & \colhead{(min)} & \colhead{} & \colhead{(km\,s$^{-1}$)} & \colhead{}
}
\startdata
RetII-409$^{\dagger}$  & 55.29047 & $-$53.81997 & 20.08 & $-3.03 \pm 0.31$ & $-2.69 \pm 0.30$ & 200 & 35 & $71.1\pm2.2$ & Likely Member (r-II) \\
RetII-406$^{\dagger}$  & 54.23542 & $-$53.74857 & 19.96 & $-1.54 \pm 0.19$ & $-1.62 \pm 0.36$ & 45 & 23 & $69.9\pm2.9$ & RV-consistent\\ 
RetII-502$^{*}$                & 54.17943 & $-53.89821$  & 19.95 & $-2.70\pm0.30$ & $-2.57\pm0.32$ & 50 & 15 & $62.4\pm9.1$ & RV-consistent\\
RetII-1                & 52.27819 & $-$54.02193 & 20.27 & $-1.91 \pm 0.25$ & $-2.40 \pm 0.27$ & 50 & 18 & $212.3:^{\ddagger}$ & Non-member \\
RetII-3                & 53.25879 & $-$53.39815 & 20.31 & $-1.84 \pm 0.24$ & $-2.12 \pm 0.22$ & 50 & 14 & $110.6:^{\ddagger}$ & Non-member\\
RetII-404              & 53.46222 & $-$54.56439 & 18.65 & $-2.53 \pm 0.20$ & $-2.49 \pm 0.28$ & 30 & 33 & $276.8\pm8.3$ & Non-member\\
RetII-5                & 53.82054 & $-$53.25934 & 19.39 & $-1.96 \pm 0.18$ & $-1.58 \pm 0.29$ & 35 & 23 & $161.0\pm2.9$ & Non-member\\
RetII-7                & 54.42821 & $-$54.37690 & 19.87 & $-1.52 \pm 0.19$ & $-1.91 \pm 0.25$ & 35 & 18 & $200.9\pm4.8$ & Non-member\\
RetII-501$^{*}$                & 52.87620 & $-53.39799$  & 20.00 & $-2.17\pm0.20$ & $-2.69\pm0.30$ & 50 & 20 & $139.8\pm4.1$ & Non-member\\ [-0.5em]
\enddata
\tablenotetext{$^{\dagger}$}{RetII-409 and RetII-406 have radial velocities consistent with membership to Reticulum II. RetII-409 is $r$-process-enhanced (see Section~\ref{sec:retii}).}
\tablenotetext{$^{\ddagger}$}{RetII-1 and RetII-3 have noisy H$\alpha$ line profiles, and therefore have highly uncertain radial velocities.}
\tablenotetext{$^{*}$}{Observed on a different date from the other candidates (see Section~\ref{sec:retii}, end of paragraph 3).}
\end{deluxetable*}

\begin{deluxetable}{lccccc}
\tablecaption{RetII-409 chemical abundances from MagE. 
\label{tab:retiiabundances}}
\tablehead{
\colhead{Star ID} & \colhead{$T_{\text{eff}}$} & \colhead{$\log\,g$} & \colhead{[C/Fe]} & \colhead{[Eu/Fe]} & \colhead{[Ba/Fe]} \\
\colhead{} & \colhead{(K)} & \colhead{(dex)} & \colhead{(dex)} & \colhead{(dex)} & \colhead{(dex)}}\\
\startdata
RetII-409 & 5476 & 3.46 & $0.55\pm0.28$  & 1.95:$^{\dagger}$ & $1.24\pm 0.28$ \\[-0.5em]
\enddata
\tablenotetext{$^{\dagger}$}{Highly uncertain due to continuum noise (see Figure~\ref{fig:retii409}).}
\end{deluxetable}

\subsection{On-sky visualizations of metal-poor substructure}
\label{sec:substructure}

The preceding subsections demonstrate the utility of MAGIC photometric metallicities in studying individual satellites and substructures in the Milky Way halo. 
Here, we present a first global exploration of the MAGIC catalog by presenting on-sky density maps of stars with low metallicities at various inferred photometric distances.

In these maps, we use RGB-assumed values for all stars (i.e., [Fe/H]$_{\text{RGB}}, d_{\text{RGB}}$), rather than restricting to those classified as RGB via \textit{Gaia} DR3 parallax to maximize completeness at large distances ($\gtrsim 70$\,kpc).
To mitigate main-sequence contaminants, we remove stars with large proper-motions in \textit{Gaia} DR3 (by requiring $|\mu_\alpha \cos\delta| < 5$\,mas\,yr$^{-1}$ and $|\mu_\delta| < 5$\,mas\,yr$^{-1}$), require [Fe/H]$_{\text{RGB}} < -2.0$ as a baseline, and require each pixel in the map to have $\geq 2$ co-moving stars at the $2\sigma$ consistency in the \textit{Gaia} DR3 proper motion catalog in order to be non-zero in these maps. 
The latter co-moving selection criterion also helps isolate coherent features relative to the ambient halo.
Additionally, we follow the quality cuts described in Section~\ref{sec:failures}, including requiring that the photometric metallicity uncertainty be $< 0.5$\,dex.

\begin{figure*}[htbp]
    \centering
    \includegraphics[width=1\textwidth]{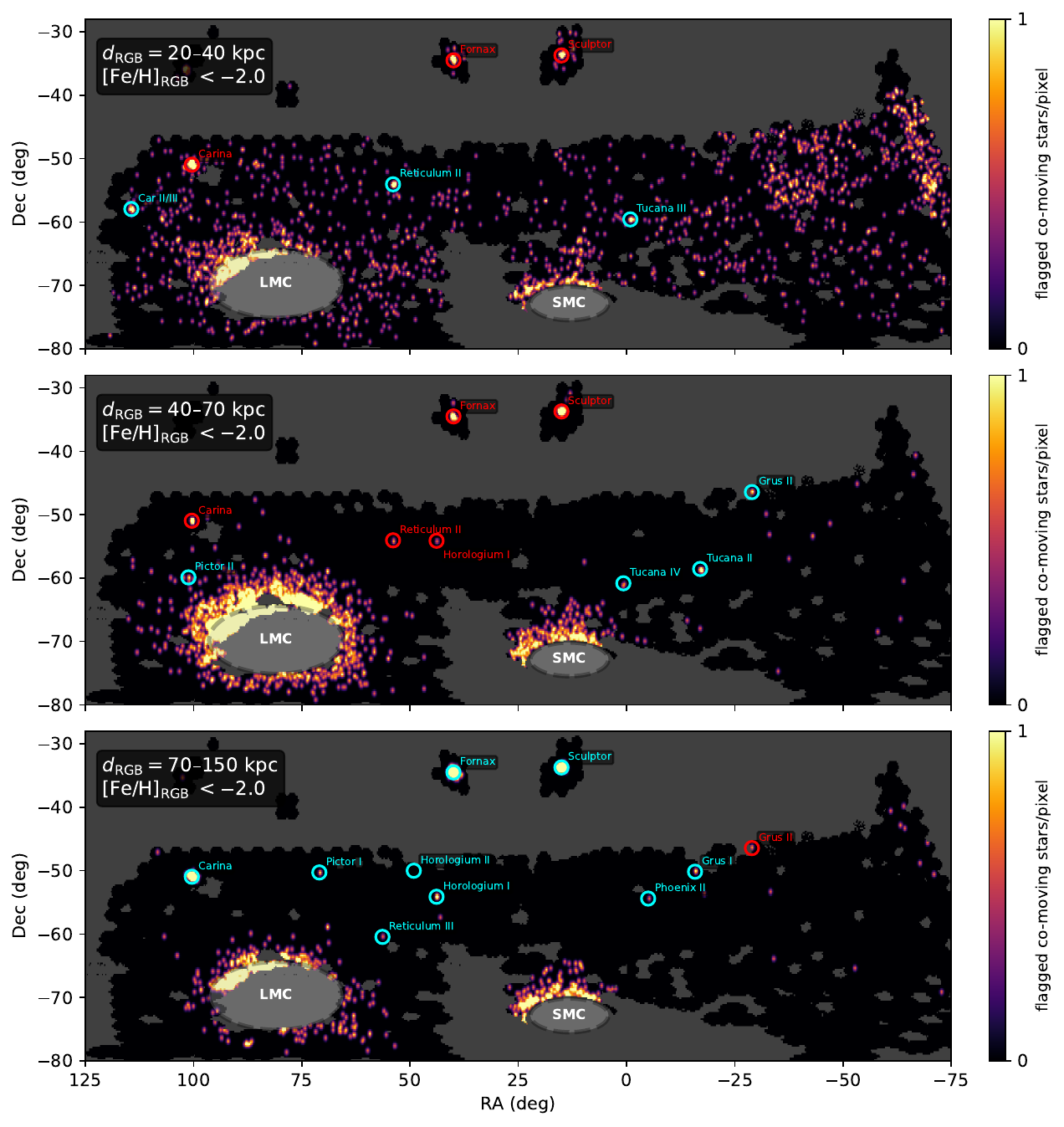}
    \caption{On-sky density maps of stars with MAGIC [Fe/H]$_{\rm RGB} < -2.0$ in three heliocentric distance bins derived from RGB-assumed photometric distances: 20--40\,kpc (top), 40--70\,kpc (middle), and 70--150\,kpc (bottom). 
    Each pixel spans $15\arcmin \times 15\arcmin$; only pixels containing $\geq 2$ stars with mutually consistent \textit{Gaia} DR3 proper motions at the $2\sigma$ level are shown. 
    The color scale indicates the smoothed number of co-moving stars per pixel, and dark gray pixels correspond to regions outside the MAGIC survey footprint. 
    Cyan circles mark known Milky Way satellite galaxies within the corresponding distance range from the Local Volume Database \citep{pace+25}, with Sculptor, Fornax, and Carina shown in red when their overdensities also appear in other distance ranges.}
    \label{fig:densityfehm2}
\end{figure*}

In Figure~\ref{fig:densityfehm2}, we present a density map of stars with [Fe/H]$_{\text{RGB}}$ $< -2.0$ following the above criteria, after pixelizing the sky in $15\arcmin \times 15\arcmin$ bins and dividing the sample into three heliocentric distance bins ($d_{\text{RGB}} = $20--40\,kpc, 40--70\,kpc, and 70--150\,kpc), using a gaussian smoothing filter with a FWHM two times the pixel size. 
Dark grey regions in the figure correspond to areas with no coverage in the MAGIC catalog, and cyan and red circles correspond to the location of known Milky Way dwarf galaxies \citep{pace+25}. 
We note several key takeaways from this map. 
First, nearly all known Milky Way dwarf galaxies (except Horologium II) in our survey footprint have an associated overdensity. 
Second, the number of ambient overdensities in the Milky Way halo in our catalog sharply drops once $d > 40$\,kpc. 
Third, we note some leakage from stars in larger classical dwarf galaxies in nearer distance bins (denoted by red circles), likely due to stars on the asymptotic giant branch being assigned RGB distances.
Overall, our map demonstrates that the MAGIC catalog can be used to recover faint overdensities in the outer Milky Way halo for future explorations. 

\begin{figure*}[htbp]
    \centering
    \includegraphics[width=1\textwidth]{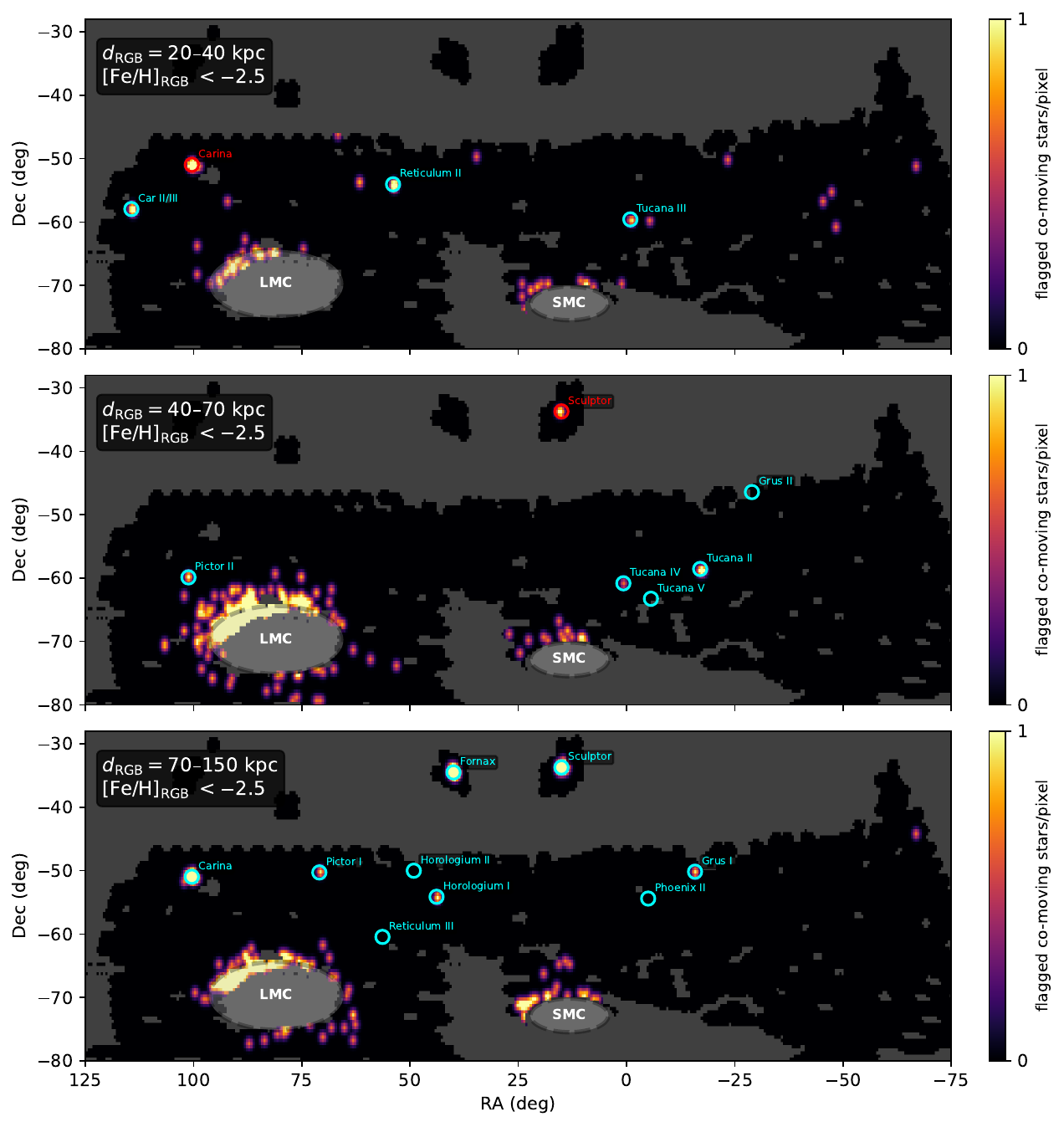}
    \caption{Same as Figure~\ref{fig:densityfehm2}, but for stars with MAGIC [Fe/H]$_{\rm RGB} < -2.5$ and with pixel size $30\arcmin \times 30\arcmin$.}
    \label{fig:densityfehm2p5}
\end{figure*}

Figure~\ref{fig:densityfehm2p5} repeats the above exercise for [Fe/H]$_{\rm RGB} < -2.5$ while expanding the pixel size to $30\arcmin \times 30\arcmin$.
This more stringent metallicity threshold reduces nearly all ambient over-densities in the Milky Way halo across distances, but still preserves a number of Milky Way satellite galaxies.
These maps illustrate the sensitivity of MAGIC in probing the low-surface-brightness regime of the lowest metallicity substructures in our Galaxy, for future explorations targeting more diffuse substructures (e.g., with the STREAMFINDER algorithm; \citealt{mi+18, imt+24}).
In Figure~\ref{fig:densityfehm2_nopm}, we show the density map of stars with [Fe/H]$_{\rm RGB} < -2.0$, without the requirement that each pixel have $\geq2$ co-moving stars, as a comparison point for the gain from proper motion information in Figures~\ref{fig:densityfehm2} and~\ref{fig:densityfehm2p5}.

\subsection{Initial EMP star follow-up}
\label{sec:initialemp}

As discussed in Section~\ref{sec:aims}, a key scientific area for MAGIC is the identification and follow-up of the lowest metallicity stars across the Milky Way, MCs, and satellite dwarf galaxies. 
Systematic spectroscopic campaigns targeting MAGIC metal-poor candidates are underway across multiple facilities, including Gemini/GMOS and Gemini/GHOST (Stringfellow et al., in prep., Skeffington et al., in prep) and Magellan/MagE and Magellan/MIKE (e.g., \citealt{plc+25, cpp+25}, Prabhu et al., in prep).
In this subsection, we present results from our very initial follow-up of candidate low-metallicity stars based on MAGIC metallicities from the medium-resolution MagE spectrograph on the 6.5\,m Magellan/Baade Telescope obtained on 2023 December 6-8, with the same observing setup that was used for the Reticulum II result in Section~\ref{sec:retii}. 

\begin{figure}[htbp]
    \centering
    \includegraphics[width=1\columnwidth]{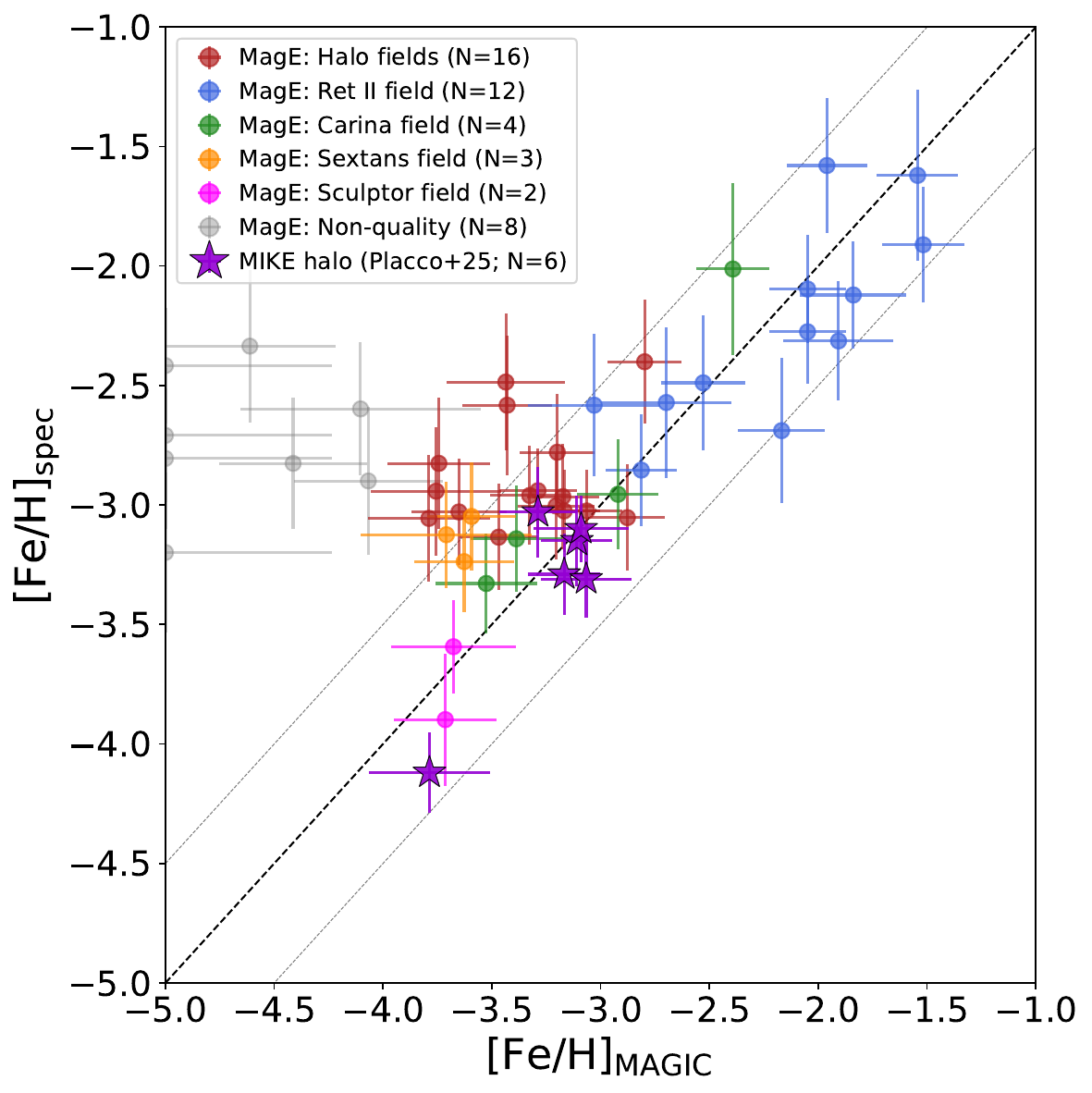}
    \caption{Outcome of the initial targeting of low-metallicity stars with the DECam MAGIC survey (see Section~\ref{sec:initialemp}). 
    We show metallicities derived from Magellan/MagE spectra (circles) for 45 candidate metal-poor stars in various regions of the sky, and with/without quality cuts in the MAGIC catalog. 
    Grey points indicate stars that fail the quality cuts described in Section~\ref{sec:failures} (predominantly due to having [Fe/H]$_{\text{MAGIC}} < -4.0$), and were targeted to explore this regime of the catalog.
    We additionally show the low-metallicity stars targeting from Magellan/MIKE in \citet{plc+25} as purple stars.
    The black dashed line denotes the one-to-one correspondence, with the grey lines denoting $\pm0.5$\,dex in metallicity. 
    Initial targeting demonstrates the recovery of low-metallicity stars with [Fe/H]$_{\text{MAGIC}}~<~-2.5$ and [Fe/H]$_{\text{MAGIC}}~<~-3.0$, while emphasizing utility of requiring [Fe/H]$_{\text{MAGIC}}~>~-4.0$ when targeting low-metallicity stars and analyzing MAGIC metallicities.}
    \label{fig:lowfehfollowup}
\end{figure}

A total of 44 sources were observed.
These sources were selected based on: (1) Their low MAGIC [Fe/H] in the version of the catalog at the time of targeting (November 2023); (2) their location in the vicinity of the Reticulum II,  Carina, and Sextans dwarf galaxies. 
In Figure~\ref{fig:lowfehfollowup}, we plot the MagE spectroscopic [Fe/H] of these stars (derived following Section~\ref{sec:retii}) versus their current MAGIC [Fe/H], and include RetII-501 and RetII-502 in this plot (see Table~\ref{tab:mageobservations}).
Stars that fail the quality cuts described in Section~\ref{sec:failures} are shown as gray datapoints in this plot, almost exclusively due to their photometric [Fe/H]$_{\text{MAGIC}} < -4.0$, and were targeted to explore the population of stars in this regime of the MAGIC catalog.
One source in the Sextans field is not plotted, as its MagE spectrum indicated it to be a non-stellar source.
We note that the stars with the two lowest MagE-based metallicities in Figure~\ref{fig:lowfehfollowup} are recoveries of two known low-metallicity members of the Sculptor dSph (Scl07-50 in \citealt{mjh+10}, and Scl031\_11 in \citealt{sht+13}).
Stars that were observed with Magellan/MIKE in \citet{plc+25} are shown as magenta stars in Figure~\ref{fig:lowfehfollowup}.

Overall, we find a reasonable recovery range of low-metallicity stars when limiting our sample to the quality cuts in Section~\ref{sec:failures} (notably, [Fe/H]$_{\text{MAGIC}} > -4.0$). 
Of the 28 stars that pass the quality cuts and are selected to have MAGIC [Fe/H] $< -2.5$, 25 (${\sim}89$\%) indeed have MagE spectroscopic [Fe/H] $< -2.5$.
Similarly, 13/22 stars (${\sim}59\%$) that pass quality cuts with MAGIC [Fe/H] $< -3.0$ have MagE spectroscopic [Fe/H] $< -3.0$.
When including stars with [Fe/H]$_{\text{MAGIC}} < -4.0$ and removing other quality flags, these fractions become 31/36 (${\sim}86$\%) for [Fe/H] $< -2.5$, and 14/30 (${\sim}47$\%) for [Fe/H] $< -3.0$. 
Future work will constrain these fractions with larger samples of low-metallicity stars, and explore the cause of the population of outliers at [Fe/H]$_{\text{MAGIC}} < -4.0$ in our catalog. 
Potential explanations include second-order CaHK magnitude zeropoint corrections (e.g., that can be addressed with UberCal; \citealt{psf+08}) or subtle imperfections in the synthetic photometry, which may disproportionately impact metallicity estimates in the most metal-poor regime.

\section{Conclusion \& Planned Data Releases}
\label{sec:conclusion}

We have presented an overview of the DECam Mapping the Ancient Galaxy in CaHK (MAGIC) survey, a 54-night NOIRLab Survey Program imaging $\gtrsim$5,000\,deg$^2$ of the southern hemisphere using a narrow-band filter covering the Ca~{\sc{ii}}~H\&K lines on the DECam/Blanco 4\,m Telescope.
We have shown that by combining DECam CaHK photometry with broadband $g,r,i$ photometry from DELVE DR2 and astrometry from \textit{Gaia} DR3, we can derive photometric metallicities for stars across our data set, enabling studies of dwarf galaxies \citep[e.g.,][]{bcl+25, cpp+25}, low-metallicity stars \citep[e.g.,][]{plc+25, cpp+25}, faint substructures \citep[e.g., ][]{apk+26}, stellar streams \citep{dcf+26}, and the Milky Way halo (see Section~\ref{sec:substructure}).
In Section~\ref{sec:validation}, we present an initial validation of our metallicities relative to those presented in APOGEE DR17 \citep{aaa+22}, \textit{Gaia} XP spectrophotometric data \citep{andrae2023}, Pristine DR1 \citep{msy+24}, and several other spectroscopic works (see Figure~\ref{fig:validation} caption). 
These comparisons show nominal scatters of $\sim$0.3\,dex down to [Fe/H] $\approx -3.0$, with slightly improved precision when [Fe/H] $< -2.0$ ($\sigma$ = 0.24\,dex) and potential systematic effects relative to certain catalogs when [Fe/H] $\gtrsim -1.0$.
Future work will improve this performance through investigating second-order photometric zero-point offsets across the dataset (e.g., Section~\ref{sec:initialemp}), and the performance at higher metallicities relative to spectroscopic metallicity measurements.

We summarize our primary scientific goals in Section~\ref{sec:aims}, which include the discovery of the lowest metallicity stars, constraining the metallicity distribution of the Milky Way and its satellites, detecting faint substructures in the Milky Way halo, and probing the faint peripheries of dwarf galaxies.
In Section~\ref{sec:science}, we show that early science results demonstrate that the MAGIC dataset already enables key results in each area.
For instance, \citet{bcl+25} leveraged MAGIC metallicities to construct the largest metallicity sample for the Sculptor dSph (${{\sim}}\,3800$ stars), revealing qualitatively new insights on its metallicity distribution and gradient.
\citet{plc+25} derived detailed chemical abundances of six distant halo stars with [Fe/H]~$< -3.0$ through high-resolution spectroscopy, including a carbon-enhanced star with [Fe/H]~$= -4.12$ at $\gtrsim 30$\,kpc, demonstrating that MAGIC targeting can recover the lowest metallicity stars in the distant Milky Way.
\citet{cpp+25} presented PicII-503, a star with the lowest iron abundance known in any ultra-faint dwarf galaxy ([Fe/H]~$< -4.63$), identified in the outskirts of the Pictor II UFD through its low MAGIC metallicity.
\citet{apk+26} used MAGIC metallicities to detect faint (${\sim} 36$\,mag\,arcsec$^{-2}$) tidal tails around the Crater II dSph. 
\citet{dcf+26} used MAGIC photometry to select a relatively clean sample of members of the Jet stream to characterize its morphology.
Furthermore, in Sections~\ref{sec:retii} to~\ref{sec:initialemp}, we presented additional early science results including the confirmation of a candidate $r$-process enhanced member of Reticulum II at $>5\,r_h$ from its center (Section~\ref{sec:retii}), an on-sky visualization of distant very metal-poor stars in the Milky Way out to ${\sim}150$\,kpc (Section~\ref{sec:substructure}), and an initial spectroscopic validation of metal-poor star targeting from MAGIC (Section~\ref{sec:initialemp}).

The first public data release of the MAGIC catalog is anticipated to occur in Fall 2026, and will include all data collected within the first year of the survey (${\sim}800$\,deg$^2$, through July 2024).
The second data release will occur one year after the survey's completion (likely Fall 2028), and will include the full footprint.
These data releases will be hosted on the NOIRLab's Astro Data Lab \citep{foe+14, nfs+20}, and will include CaHK photometry along with value-added products such as photometric metallicities, photometric distances, and associated quality flags for a subset of vetted stellar sources.
The associated CaHK catalog for the early MAGIC results on Sculptor, Crater II, and the Jet stream are publicly available on Zenodo \citep{bcl+25,apk+26,dcf+26}\footnote{Sculptor dSph-- \url{https://doi.org/10.5281/zenodo.15132836};\\ Crater II dSph-- \url{https://doi.org/10.5281/zenodo.16804863};\\ 
Jet stream-- \url{https://doi.org/10.5281/zenodo.19475550}}.
We also intend to publish outcomes from ongoing spectroscopic campaigns with Gemini/GMOS, Gemini/GHOST, Magellan/MagE, and Magellan/MIKE that are characterizing the low-metallicity tail of the catalog, following on previously published results \citep{plc+25, cpp+25}, and ongoing efforts in preparation (Stringfellow et al. in prep., Skeffington et al. in prep., Prabhu et al. in prep.).

We anticipate additional scientific utility from the MAGIC dataset as it completes its proposed ${\sim}5{,}000$\,deg$^2$ coverage. 
The MAGIC data releases will strongly complement deep $u,g,r,i,z,Y$ photometry from Rubin LSST \citep{lsst}, deeper proper motions from \textit{Gaia} DR4 and onwards \citep{gaiadr4}, and ongoing and forthcoming multiplexed spectroscopic surveys (i.e., 4MOST, Subaru-PFS, DESI, WEAVE, and Spec-S5; \citealt{4most, tts+16, desi, weave, specs5}).
The discovery potential in the local Universe remains rich, and the MAGIC survey provides a crucial step in enabling the wide-field characterization of ancient, faint stellar populations across the Milky Way and its satellite galaxies.

\textit{Facilities:} CTIO:Blanco (DECam), Magellan:Baade (MagE)

\textit{Software:} Astropy \citep{astropy, astropy2}, NumPy \citep{numpy}, SciPy \citep{scipy}, Matplotlib \citep{Hunter+07}, SMHR \citep{c+14}, MOOG \citep{s+73}, alexmods (https://github.com/alexji/alexmods), MagE CarPy reduction pipeline \citep{k+03}.

\section*{Acknowledgments}

A.C. is supported by the Brinson Foundation through a Brinson Prize Fellowship grant.
The work of V.M.P., C.E.M-V, Y.C., and K.A.G.O. is supported by NOIRLab, which is managed by the Association of Universities for Research in Astronomy (AURA) under a cooperative agreement with the U.S. National Science Foundation.
DJS acknowledges support from NSF grant AST-2508746.
W.C. gratefully acknowledges support from a Gruber Science Fellowship at Yale University. This material is based upon work supported by the National Science Foundation Graduate Research Fellowship Program under Grant No. DGE2139841. Any opinions, findings, and conclusions or recommendations expressed in this material are those of the author(s) and do not necessarily reflect the views of the National Science Foundation.
J.S. acknowledges financial support from PID2022-138896NB-C53 and the Severo Ochoa grant CEX2021-001131-S funded by MCIN/AEI/ 10.13039/501100011033.
D.L.N. acknowledges funding from National Science Foundation grant AST-2408159.
B.M.P. acknowledges support from NSF grant AST-2508745.
D.C. acknowledges support from NSF grant AST-2508747.
This work was completed in part with resources provided by the University of Chicago's Research Computing Center.
We thank Oleg Gnedin and Eric Bell for helping coordinate the MagE observation of two of the Reticulum II candidate members on 2023 October 14.

The DELVE project is partially supported by the NASA Fermi Guest Investigator Program Cycle 9 No. 91201.
This work is partially supported by Fermilab LDRD project L2019-011. 
This material is based upon work supported by the National Science Foundation under Grant No. AST-2108168, AST-2108169, AST-2307126, and AST-2407526.

This project used data obtained with the Dark Energy Camera (DECam), which was constructed by the Dark Energy Survey (DES) collaboration. 
Funding for the DES Projects has been provided by the US Department of Energy, the U.S. National Science Foundation, the Ministry of Science and Education of Spain, the Science and Technology Facilities Council of the United Kingdom, the Higher Education Funding Council for England, the National Center for Supercomputing Applications at the University of Illinois at Urbana--Champaign, the Kavli Institute for Cosmological Physics at the University of Chicago, the Center for Cosmology and Astro-Particle Physics at the Ohio State University, the Mitchell Institute for Fundamental Physics and Astronomy at Texas A\&M University, Financiadora de Estudos e Projetos, Funda\c{c}\~{a}o Carlos Chagas Filho de Amparo \`{a} Pesquisa do Estado do Rio de Janeiro, Conselho Nacional de Desenvolvimento Cient\'{i}fico e Tecnol\'{o}gico and the Minist\'{e}rio da Ci\^{e}ncia, Tecnologia e Inova\c{c}\~{a}o, the Deutsche Forschungsgemeinschaft and the Collaborating Institutions in the Dark Energy Survey.

The Collaborating Institutions are Argonne National Laboratory, the University of California at Santa Cruz, the University of Cambridge, Centro de Investigaciones En\'{e}rgeticas, Medioambientales y Tecnol\'{o}gicas--Madrid, the University of Chicago, University College London, the DES-Brazil Consortium, the University of Edinburgh, the Eidgen\"{o}ssische Technische Hochschule (ETH) Z\"{u}rich, Fermi National Accelerator Laboratory, the University of Illinois at Urbana-Champaign, the Institut de Ci\`{e}ncies de l'Espai (IEEC/CSIC), the Institut de F\'{i}sica d'Altes Energies, Lawrence Berkeley National Laboratory, the Ludwig-Maximilians Universit\"{a}t M\"{u}nchen and the associated Excellence Cluster Universe, the University of Michigan, NSF NOIRLab, the University of Nottingham, the Ohio State University, the OzDES Membership Consortium, the University of Pennsylvania, the University of Portsmouth, SLAC National Accelerator Laboratory, Stanford University, the University of Sussex, and Texas A\&M University.

Based on observations at NSF Cerro Tololo Inter-American Observatory, NSF NOIRLab (NOIRLab Prop. ID 2019A-0305; PI: Drlica-Wagner, NOIRLab Prop. ID 2023B-646244; PI: Anirudh Chiti, NOIRLab Prop. ID 2023A-933926; P.I. Chiti, NOIRLab Prop. ID 2024A-167177; P.I. Cerny, NOIRLab Prop. ID 2024A-930400; P.I. Cerny, NOIRLab Prop. ID 2024A-974884; P.I. Pace, NOIRLab Prop. ID 2024B-255918; P.I. Pace, NOIRLab Prop. ID 2024B-616175; P.I. Cerny, NOIRLab Prop. ID 2025A-303722; P.I. Carballo-Bello \& Chiti, NOIRLab Prop. ID 2025A-402104; P.I. Pace, NOIRLab Prop. ID 2025A-568024; P.I. Do \& Chiti, NOIRLab Prop. ID 2025A-942172; P.I. Mart\'inez-V\'azquez, 2025B-444868; P.I. Pace), which is managed by the Association of Universities for Research in Astronomy (AURA) under a cooperative agreement with the U.S. National Science Foundation.

This manuscript has been authored by Fermi Research Alliance, LLC under Contract No. DE-AC02-07CH11359 with the U.S. Department of Energy, Office of Science, Office of High Energy Physics. The United States Government retains and the publisher, by accepting the article for publication, acknowledges that the United States Government retains a non-exclusive, paid-up, irrevocable, world-wide license to publish or reproduce the published form of this manuscript, or allow others to do so, for United States Government purposes.

This work has made use of data from the European Space Agency (ESA) mission {\it Gaia} (\url{https://www.cosmos.esa.int/gaia}), processed by the {\it Gaia} Data Processing and Analysis Consortium (DPAC, \url{https://www.cosmos.esa.int/web/gaia/dpac/consortium}). Funding for the DPAC has been provided by national institutions, in particular the institutions participating in the {\it Gaia} Multilateral Agreement.

This research uses services or data provided by the Astro Data Archive at NSF's NOIRLab. NOIRLab is operated by the Association of Universities for Research in Astronomy (AURA), Inc. under a cooperative agreement with the National Science Foundation.

\bibliography{bibliography}{}
\bibliographystyle{aasjournal}

\appendix

\section{Affiliated DECam CaHK Programs}
\label{app:affiliated}

Table~\ref{tab:affiliated} lists the affiliated P.I.\ programs that have obtained DECam CaHK imaging and processed with the MAGIC survey pipeline, as described in
Section~\ref{sec:strategy}.

\begin{deluxetable}{lccc}
\tablecaption{Affiliated DECam CaHK observing programs.\label{tab:affiliated}}
\tablehead{
\colhead{NOIRLab Prop.\ ID} & \colhead{P.I.(s)} & \colhead{Semester} & \colhead{Data Obtained}
}
\startdata
2023A-933926 & Chiti              & 2023A & Carina dSph, Sextans dSph, Reticulum II UFD, Jet stream spur,  \\
& & & NGC~1904, NGC~2808, NGC~5024, NGC~5053, NGC~5824, \\
& & & NGC~5897, NGC~5927\\
2024A-167177 & Cerny              & 2024A & Bootes~I, Centaurus I, Carina II, Carina~III, Hydra~II, Leo~IV \\
2024A-930400 & Cerny              & 2024A & DELVE~1 \\
2024A-974884 & Pace               & 2024A & Outskirts of Sextans and Crater II stream \\
2024B-255918 & Pace               & 2024B & Large tiling of Sculptor and Fornax \\
2024B-616175 & Cerny              & 2024B & Balbinot 1, Kim 1 \\
2025A-303722 & Carballo-Bello, Chiti & 2025A & NGC 1851, NGC~2298, NGC~288, NGC~4372, NGC~4590, \\
& & & NGC~5897, NGC~5986, NGC~6254, NGC~6397, NGC~6752, \\
& & & NGC~6981, NGC~7089, NGC~7099, Rup~106, Whiting~1\\
2025A-402104 & Pace               & 2025A & Crater II stream, Hercules, DELVE 1 region  \\
2025A-568024 & Do, Chiti          & 2025A & Jet stream \\
2025A-942172 & Mart\'inez-V\'azquez & 2025A & Omega Cen \& Fimbulthul stream \\
2025B-444868 & Pace               & 2025B & Outskirts of Fornax, Hydrus I, and Antlia II \\[-0.5em]
\enddata
\tablecomments{These programs account for 20.5 allocated nights of DECam CaHK imaging, in addition to the 54-night MAGIC survey program (NOIRLab Prop.\ ID
2023B-646244; P.I.\ Chiti).}
\end{deluxetable}

\section{Dwarf Galaxy Samples in Figure~9}
\label{app:dwarfxmatch}

The left panel of Figure~\ref{fig:validation} includes spectroscopic metallicities from a number of dwarf galaxy studies cross-matched with the MAGIC catalog.
Note that this comparison only includes stars that can be classified as RGB stars based on Section~\ref{sec:derivationdistances}. 
Accordingly, the cross-matched samples to individual studies are small due to the distances to these systems (generally $\gtrsim30$\,kpc; \citealt{pace+25}). 
Here we list each study, the corresponding dwarf galaxies, and the number of cross-matched stars passing the quality cuts described in Section~\ref{sec:failures} (905 stars total):
\citet{sdl+15}-- Reticulum~II (4 stars);
\citet{koposov+15}-- Horologium~I (3 stars);
\citet{li+2018carina}-- Carina~II and Carina~III (4 stars);
\citet{cfj+18}-- Tucana~II (6 stars);
\citet{lsk+18}-- Tucana~III (7 stars);
\citet{fcb+19}-- Phoenix~II (1 star);
\citet{sle+20}-- Grus~II, Tucana~IV (2 stars);
\citet{csf+22}-- Grus~I (2 stars);
\citet{chiti2023}-- Tucana~II (3 stars);
\citet{tsa+23}-- Sculptor (502 stars);
\citet{cml+24}-- LMC (8 stars);
\citet{oh+24}-- LMC (5 stars);
\citet{ljs+24}-- Fornax and Carina (3 stars);
\citet{hlp+24}-- Centaurus~I and Eridanus~IV (6 stars);
\citet{plj+25}-- Pictor~II (1 star);
\citet{tba+25}-- Sextans (138 stars);
and \citet{gpp+26}-- Eridanus~IV, Fornax, Sculptor, Sextans (210 stars).

\section{Density Map Without Proper Motion Filtering}
\label{sec:appendixdensity}
\begin{figure*}[htbp]
    \centering
    \includegraphics[width=1\textwidth]{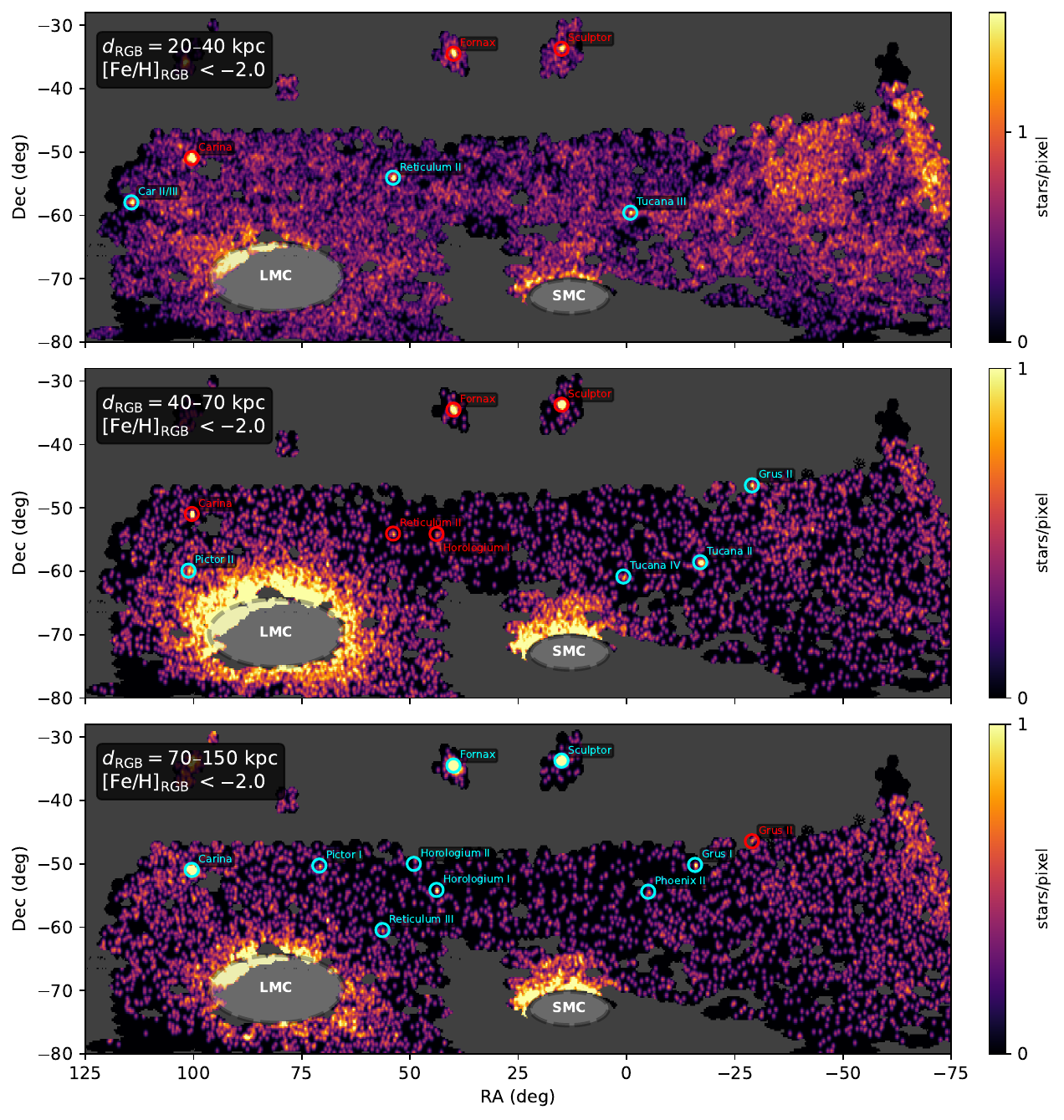}
    \caption{Same as Figure~\ref{fig:densityfehm2}, but without the requirement that a pixel have $\geq$2 co-moving stars in \textit{Gaia} DR3 \citep{gaiadr3} to have color. The lack of sources in the bottom left of the density map is due to the requirement that $E(B-V) < 0.2$ in \citet{sfd+98} (see Section~\ref{sec:failures}).}
    \label{fig:densityfehm2_nopm}
\end{figure*}

In Figure~\ref{fig:densityfehm2_nopm}, we show an on-sky density map of stars with [Fe/H]$_{\rm RGB} < -2.0$ without the proper motion consistency filter applied in Figure~\ref{fig:densityfehm2}. 
This illustrates the gain from using \textit{Gaia} DR3 proper motions to isolate pixels with co-moving stars.

\end{document}